\documentclass[12pt,aps,prb,tightenlines,showpacs]{revtex4}
\usepackage{times}
\usepackage[T1]{fontenc}
\usepackage[latin1]{inputenc}
\usepackage{amsmath}
\usepackage{graphics}
\usepackage{amssymb}
\usepackage{subfigure}

\makeatletter

\providecommand{\LyX}{L\kern-.1667em\lower.25em\hbox{Y}\kern-.125emX\@}

\usepackage{times}
\usepackage[T1]{fontenc}
\usepackage[latin1]{inputenc}
\usepackage{amsmath}
\usepackage{graphics}
\usepackage{amssymb}
\usepackage{subfigure}
\DeclareMathOperator{\sech}{sech}

\makeatother

\begin{document}

\pacs{75.10.Hk, 75.25.+z, 75.40.Cx}

\title{Critical behavior of 2 and 3 dimensional ferro- and antiferromagnetic spin
ice  systems in the framework of the Effective Field Renormalization Group technique}

\author{Angel J. Garcia-Adeva}

\email{garcia@landau.physics.wisc.edu}

\author{David L. Huber}

\affiliation{Department of Physics; University of Wisconsin--Madison; Madison, WI 53706}

\begin{abstract}
\bigskip{}
In this work we generalize and subsequently apply the Effective Field Renormalization
Group technique to the problem of ferro- and antiferromagnetically coupled Ising
spins with local anisotropy axes in geometrically frustrated geometries (\emph{kagomé}
and pyrochlore lattices). In this framework, we calculate the various ground
states of these systems and the corresponding critical points. Excellent agreement
is found with exact and Monte Carlo results. The effects of frustration are
discussed. As pointed out by other authors, it turns out that the spin ice model
can be exactly mapped to the standard Ising model but with effective interactions
of the opposite sign to those in the original Hamiltonian. Therefore, the ferromagnetic
spin ice is frustrated, and does not order. Antiferromagnetic spin ice (in both
2 and 3 dimensions), is found to undergo a transition to a long range ordered
state. The thermal and magnetic critical exponents for this transition are calculated.
It is found that the thermal exponent is that of the Ising universality class,
whereas the magnetic critical exponent is different, as expected from the fact
that the Zeeman term has a different symmetry in these systems. In addition,
the recently introduced Generalized Constant Coupling method is also applied
to the calculation of the critical points and ground state configurations. Again,
a very good agreement is found with both exact, Monte Carlo, and renormalization
group calculations for the critical points. Incidentally, we show that the generalized
constant coupling approach can be regarded as the lowest order limit of the
EFRG technique, in which correlations outside a frustrated unit are neglected,
and scaling is substituted by strict equality of the thermodynamic quantities.
\end{abstract}
\maketitle

\section{Introduction}

\begin{figure}
{\par\centering \subfigure[\emph{Kagom\'{e}} lattice.]{\resizebox*{0.45\textwidth}{!}{\includegraphics{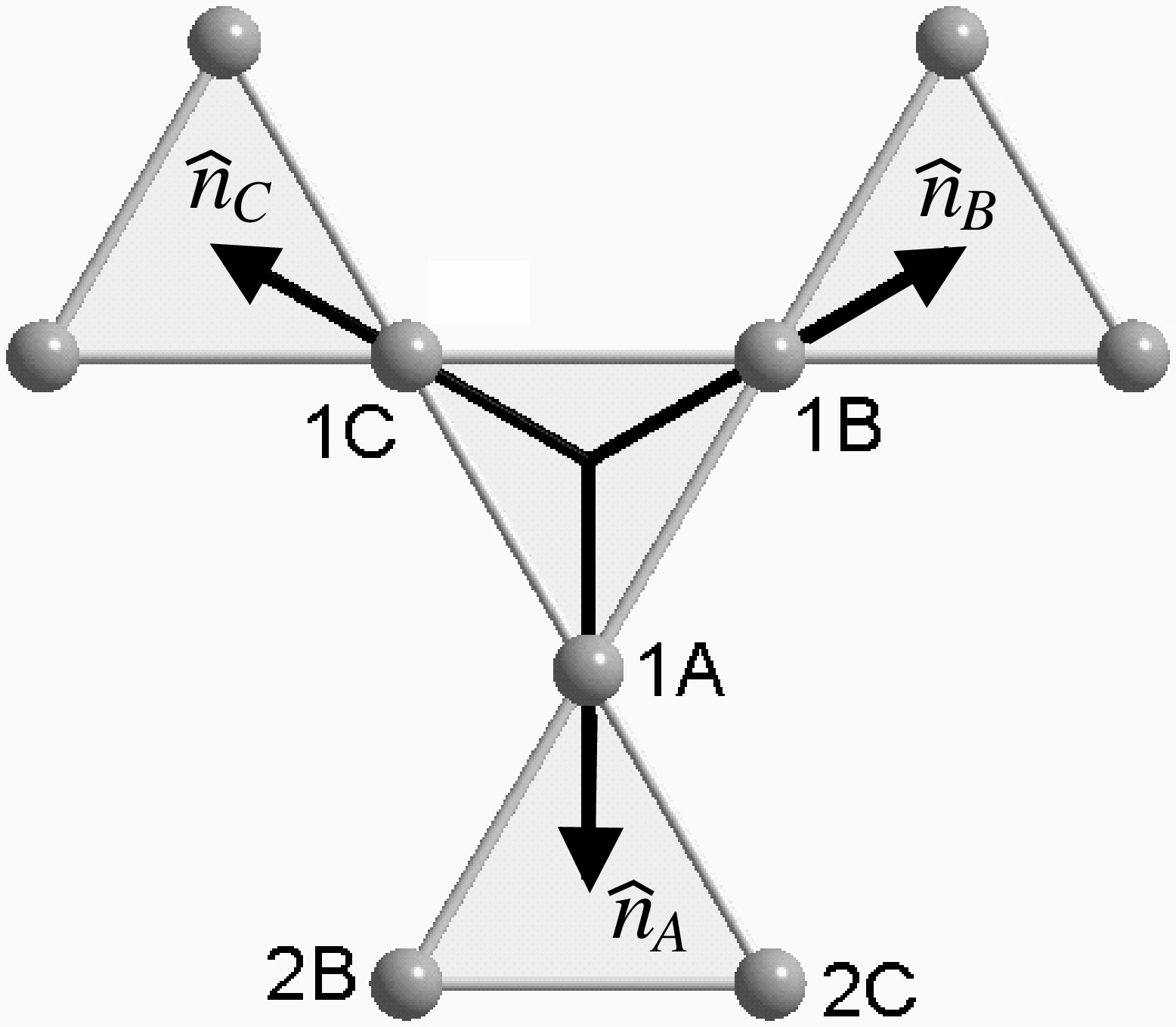}}} \subfigure[Pyrochlore lattice.]{\resizebox*{0.45\textwidth}{!}{\includegraphics{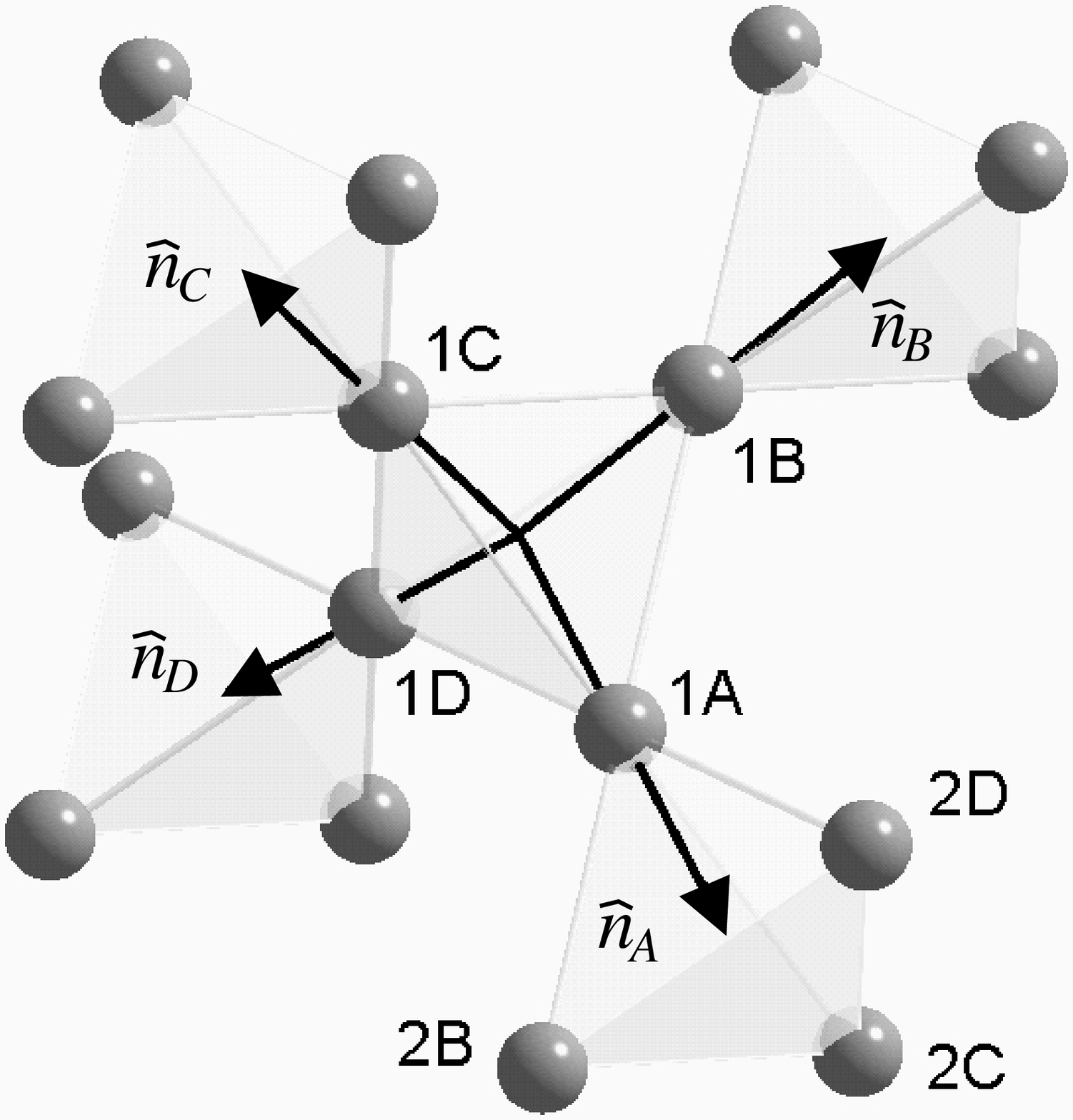}}} \par}

\caption{\label{fig.structure}The structures considered in this work. The black thick
lines indicate the axes of anisotropy. Spin labels are of the form \protect\protect\( i\alpha \protect \protect \),
where \protect\protect\( \alpha \protect \protect \) is the label of the sublattice
and \protect\protect\( i\protect \protect \) the index of the spin inside the
corresponding sublattice. The normal vectors \protect\( \widehat{n}_{\alpha }\protect \)
have been also depicted.}
\end{figure}

Geometrically frustrated antiferromagnets (GFAF) have emerged in the last years
as a new class of magnetic materials with many unusual physical properties\cite{GINGRAS2000,Ramirez1994,SCHIFFER1996b,MOESSNER2000}.
In these materials, the elementary magnetic unit is the triangle, which makes
it impossible to satisfy all the antiferromagnetic bonds at the same time, with
the result of a macroscopically degenerate ground state. Examples of GFAF are
the pyrochlore and the \emph{kagomé} lattices. In the former, the magnetic ions
occupy the corners of a 3D arrangement of corner sharing tetrahedra; in the
later, the magnetic ions occupy the corners of a 2D arrangement of corner sharing
triangles (see Fig. \ref{fig.structure}). In the case of materials which crystallize
in the pyrochlore structure, the static magnetic susceptibility follows the
Curie--Weiss law down to temperatures well below the Curie--Weiss temperature.
At this point, usually one to two orders of magnitude smaller than the Neél
point predicted by the standard mean field (MF) theory, some systems exhibit
some kind of long range order (LRO), whereas others show a transition to a spin
glass state (SG). This is a striking feature for a system with only a marginal
amount of disorder. Finally, there are some pyrochlores which do not exhibit
any form of order whatsoever, and are usually regarded as spin liquids. In the
case of the \emph{kagomé} lattice, even though there are very few real systems
where this structure is realized, the magnetic properties fall in two major
categories: the vast majority of the compounds studied show a transition to
a LRO state with a non collinear configuration of spins, and a few systems exhibit
no LRO, but a SG like transition.

Surprisingly, Harris and coworkers \cite{HARRIS1997,HARRIS1998,BRAMWELL1998}
recently showed that geometrical frustration can also arise from ferromagnetic
interactions. This cannot occur for Heisenberg spins but, in some cases, a strong
single ion anisotropy along high symmetry directions of the crystallographic
structure can force the spins to point in such directions, so they can effectively
be considered as Ising spins\cite{Moessner98b}. In this case, the Ising spins
with ferromagnetic interactions do not order at any finite temperature. They
analyzed this problem in the context of Ising spins in the corners of a pyrochlore
lattice, and termed this system spin ice pyrochlore, because it can be mapped
to the well know proton ordering problem in the ice lattice. Furthermore, in
full equivalence with the ice problem, the spin ice model presents a slow dynamic,
due to the existence of large energy barriers associated with rearrangements
of the spins in the ground state, resulting in a spin freezing which resembles
a spin glass, \emph{but occurring in a system where chemical disorder is absent}\cite{HARRIS1997,RAMIREZ2000}.

While at first sight this can look like a very interesting academic problem,
it turned out to be also of enormous practical interest, because various pyrochlore
oxides were found to have strong single ion anisotropy along the \( \left\langle 1\, 1\, 1\right\rangle  \)-directions
of the lattice. This effect usually occurs for rare earth magnetic ions with
a large orbital moment, as it is the case of the pyrochlores Ho\( _{2} \)Ti\( _{2} \)O\( _{7} \),
Dy\( _{2} \)Ti\( _{2} \)O\( _{7} \), and Yb\( _{2} \)Ti\( _{2} \)O\( _{7} \),
among others, which are now known to be very good realizations of the spin ice
model.

From the theoretical point of view, apart from the interest of the spin ice
system itself, this system also offers a unique opportunity for testing models
which incorporate geometrical frustration, for the obvious reason that the Ising
Hamiltonian is simpler to study mathematically than the Heisenberg one and,
in some cases, analytical results are easily obtained, which can be readily
compared with experimental data or simulations. If such a comparison is adequate,
one can get some confidence that these techniques can also be applied to the
Heisenberg problem in the pyrochlore lattice, which is an even more interesting
problem.

This is precisely the purpose of the present work. We will study the critical
properties of geometrically frustrated Ising \emph{ferro-} and \emph{antiferromagnets}
by introducing a generalization of a well known real space renormalization group
technique, the so called Effective Field Renormalization Group (EFRG) method\cite{LI1988,FITTIPALDI1994},
which has been successfully applied to a large variety of systems (see Ref.~\onlinecite{PLASCAK1999}
and references therein), and it is known to provide very good estimates of the
critical temperature and the critical exponents. We apply this technique to
two model systems: the spin ice problem in the pyrochlore lattice, and a 2D
analog to the spin ice problem, which consists of Ising spins in the \emph{kagomé}
lattice with local axes of anisotropy along the heights of the triangle. This
second example, though academic, is a very instructive one, as it will serve
as an example to develop the present method. Our first task will be to generalize
the EFRG technique to include the effect of geometrical frustration, that is,
the non-Bravais character of the lattices considered. We will then study the
existence of critical points and their values for both ferro- and antiferromagnetic
interactions. The obtained values are compared with exact values known for the
Ising model in the \emph{kagomé} lattice \cite{SYOZI1972} and with numerical
estimations of this quantity obtained by Harris and Bramwell for the spin ice
pyrochlore\cite{BRAMWELL1998}. We find that, even at the lowest order of approximation,
in which correlations with ions further that NN are neglected, the estimates
are excellent for both lattices. Furthermore, we show that these spin ice-like
Ising models can be related to the standard Ising model\footnote{%
It is important to stress that this single axis Ising spin system cannot occur
in a pyrochlore lattice, as it is incompatible with the cubic symmetry of the
spatial group. However, it is a very instructive example, and will prove to
be useful for testing the accuracy of the results presented in this work.
}. Therefore, we are able to understand why the ground state given by the ice
rules is never found in these systems, in agreement with numerical evidences
provided by other authors\cite{BRAMWELL1998,RAMIREZ2000}. Next, we evaluate
the thermal and magnetic exponents for the transition to an ordered state that
is found for antiferromagnetic spin ice, namely, a ground state in which all
the spins point out or all the spins point in. These exponents are calculated
for two limiting cases: the one corresponding to Ising spins with a single anisotropy
axis, that is, the standard Ising model, and the one corresponding to the spin
problem. For the standard Ising model the calculated exponents are found to
be in good agreement when compared with exact values for 2D lattices, and with
Monte Carlo simulations for the 3D lattice, for the Ising universality class.
For the spin ice case, the thermal exponent is the same as in the standard Ising
model, whereas the magnetic one is found to be different for both lattices,
in agreement with the fact that the Zeeman term in this case does not have the
same symmetry as in the standard Ising model.

Additionally, we find interesting to compare the results obtained from this
new method with the ones predicted by the Generalized Constant Coupling (GCC)
model, recently developed by the present authors \cite{GARCIA-ADEVA2000a,GARCIA-ADEVA2000b},
which has been shown to provide very accurate results for the magnetic properties
of Heisenberg GFAF. The results for the critical points and ground state configurations
obtained in the GCC framework are found to be in excellent agreement with the
ones found by using the EFRG technique, in spite of the mathematical simplicity
of the GCC method. Incidentally, we show that this agreement is not a coincidence,
but it is related to the fact that the GCC method can be considered as the zero-th
order approximation of the EFRG scheme.

The remainder of this paper is organized as follows: in the next section, we
introduce the systems we will deal with and a brief summary of the EFRG method,
together with the corresponding generalization to deal with geometrically frustrated
lattices. In the third section we develop the EFRG scheme for both the \emph{kagomé}
and pyrochlore lattice with anisotropic Ising interactions. In section \ref{sec.critical.behavior},
the finite scaling hypothesis is used to calculate the non-trivial fixed points
of the RG transformation (the critical points), together with the possible ground
state configurations of the system. In section \ref{sec.critical.exponents}
we calculate the thermal and magnetic critical exponents for the long range
ordered state found in the previous section for antiferromagnetic interactions.
In section \ref{sec.gcc.method}, we apply the GCC model to the present problem
and, again, the critical points and ground state configurations are calculated
and compared with the EFRG method. In section \ref{sec.connection}, the connection
between both schemes is furnished. Finally, section \ref{sec.conclusions} is
devoted to the conclusions.

\section{Statement of the problem}

In the systems\label{sec.model} we will consider below, the strong single-ion
anisotropy forces the spins in the lattice to point in high symmetry directions
and, to a first approximation, they can be regarded as Ising spins\cite{Moessner98b}.
For example, in the spin ice pyrochlore, each ion in the elementary tetrahedral
unit points in a different \( \left\langle 111\right\rangle  \)-type direction,
which connects the spin with the center of the tetrahedra. Therefore, the spin
variables are not scalar ones, but have vectorial character. We will represent
each of the spins by a vector \( \vec{s}_{i\alpha }=s_{i\alpha }\widehat{n}_{\alpha } \),
where \( s_{i\alpha } \) takes the values \( \pm 1 \) and \( \widehat{n}_{\alpha } \)
stands for the unitary vector in one of the directions of anisotropy, where
\( \alpha  \) takes the values \( \alpha =A,B,C,\ldots  \), depending on the
number of different directions of anisotropy (see Fig.~\ref{fig.structure}).
The index \( i \) runs over all the spins with a given normal vector.

The Ising Hamiltonian with only NN interactions in adimensional units can be
then put as\begin{equation}
\mathcal{H}=K\sum _{\alpha \neq \beta }\sum _{\left\langle i,j\right\rangle }\vec{s}_{i\alpha }\cdot \vec{s}_{j\beta },
\end{equation}
 where \( K=\beta J \) is the adimensional coupling constant, which is positive
for ferromagnetic interactions and negative for antiferromagnetic interactions.
We can see that, in this way, we have defined a finite number of sublattices,
characterized by the corresponding orientation of the Ising spins, which has
the advantage that a spin in sublattice \( \alpha  \) interacts only with spins
in different sublattices.

The geometries we will consider in this work are those depicted in Fig.~\ref{fig.structure}:
anisotropic Ising spins in the \emph{kagomé} lattice, which can be considered
as a 2D analog to the spin ice problem, and the spin ice problem, that is, anisotropic
Ising spins in a pyrochlore lattice.

In order to study the critical properties of these systems, we will make use
of the so called EFRG technique, which has been proved to be a very good RG
scheme for dealing with critical phenomena. We will not present the method here,
as there is an excellent review on this subject\cite{PLASCAK1999}, but only
the most important notions.

According to the idea of scaling\cite{GOLDENFELD1994}, close to the transition
the singular part of the free energy scales as\begin{equation}
f_{s}(\epsilon ,H)=l^{-d}f_{s}(l^{1/\nu }\epsilon ,l^{y_{H}}H),
\end{equation}
 where \( \epsilon =K-K_{c} \), with \( K_{c} \) the critical coupling, \( l \)
is an arbitrary scaling factor, \( d \) is the dimensionality of the lattice,
\( y_{H} \) is the magnetic critical exponent and \( H \) the applied magnetic
field, and \( \nu  \) is the correlation length critical exponent. Any other
thermodynamic quantity, \( P \), can be evaluated as derivatives of this singular
part of the free energy near the transition, and will behave as a power law\begin{equation}
P\sim \left| \epsilon \right| ^{-\sigma },
\end{equation}
 where \( \sigma  \) is the critical exponent of the quantity \( P \). For
example, the magnetization, near criticality, behaves as\begin{equation}
\label{scaling}
m(\epsilon ,H)=l^{-d+y_{H}}m(l^{1/\nu }\epsilon ,l^{y_{H}}H).
\end{equation}

On the other hand, according to the finite size scaling hypothesis, the generalized
scaling relation obeyed by any thermodynamic quantity \( P \), taking into
account the finite size \( L \) of the system, can be put as\cite{GOLDENFELD1994}\begin{equation}
\label{finite.scaling}
P(\epsilon ,H,L)=l^{\phi }P(l^{1/\nu }\epsilon ,l^{y_{H}}H,l^{-1}L),
\end{equation}
 where \( \phi =\sigma /\nu  \) is the anomalous dimension of the quantity
\( P \). By choosing a smaller system of size \( L' \), fixing the scaling
factor to the value \( l=\frac{L}{L'} \), and introducing the variables \( \epsilon '=l^{1/\nu }\epsilon =K'-K_{c} \)
and \( H'=l^{y_{H}}H \), we can rewrite relation \eqref{finite.scaling} in
the form\begin{equation}
\label{finite.scaling2}
\frac{P_{L'}(K',H')}{{L'}^{\phi }}=\frac{P_{L}(K,H)}{L^{\phi }}.
\end{equation}
 This is the basic equation from where a mapping \( (K,H)\to (K',H') \) can
be established, and the corresponding critical points and exponents obtained.

By applying relation \eqref{finite.scaling2} to different thermodynamic quantities,
different RG approaches are generated. For example, on applying \eqref{finite.scaling2}
to the correlation length, we have\begin{equation}
\label{correlation.length}
\frac{\xi '(K',H')}{L'}=\frac{\xi (K,H)}{L},
\end{equation}
 which means that, for the correlation length, the anomalous dimension is the
unity. 

The EFRG approach corresponds to apply relation \eqref{finite.scaling2} to the
order parameter (magnetization). The order parameter for two clusters of different
size is calculated by using the Callen--Suzuki identity\cite{CALLEN1963,SUZUKI1965}

\begin{equation}
\label{EFRG.main.eq}
\left\langle O_{p}\right\rangle =\left\langle \frac{Tr_{p}O_{p}e^{\mathcal{H}_{p}}}{Tr_{p}e^{\mathcal{H}_{p}}}\right\rangle _{\mathcal{H}},
\end{equation}
 where the partial trace is taken over the set of \( p \) spin variables specified
by the finite size cluster Hamiltonian \( \mathcal{H}_{p} \); \( O_{p} \)
is the corresponding order parameter, and \( \left\langle \ldots \right\rangle _{\mathcal{H}} \)
indicates the usual canonical thermal average over the ensemble defined by the
complete Hamiltonian \( \mathcal{H} \). The idea is then to replace the effect
of the total Hamiltonian not included in the partial trace in \eqref{EFRG.main.eq}
by fixed values of the magnetization of the ions outside the cluster, which
act as a symmetry breaking field, \( b \). In this way, the order parameter
in the finite cluster can be computed for \( b\ll 1 \) and \( H\ll 1 \), and
an equation of the form\begin{equation}
m_{p}(K,b,H)=\left\langle O_{p}\right\rangle =f_{p}(K)\, b+g_{p}(K)\, H,
\end{equation}
 is obtained. Repeating the same calculation for a cluster of different size,
\( p' \), and assuming that both magnetizations are related by the scaling
relation \eqref{scaling} near criticality, we obtain the following equation\begin{equation}
f_{p'}(K')\, b'+g_{p'}(K')\, H'=l^{d-y_{H}}\, f_{p}(K)\, b+l^{d-y_{H}}\, g_{p}(K)\, H.
\end{equation}
 As \( b \) and \( b' \) are, in some sense, magnetizations, they are also
assumed to satisfy \eqref{scaling}, so we have the following relation\begin{equation}
f_{p'}(K')=f_{p}(K),
\end{equation}
 from which the fixed points of this RG transformation and, thus the critical
points, are found by solving this equation for \( K'=K=K_{c} \). The corresponding
thermal eigenvalue is obtained as the corresponding eigenvalue of this recursion
relation, linearized near the critical point\begin{equation}
\label{thermal.eigenvalue}
\lambda _{T}=\left. \frac{\partial f_{p}}{\partial K}\left( \frac{\partial f_{p'}}{\partial K'}\right) ^{-1}\right| _{K_{c}},
\end{equation}
 and by making use of \eqref{correlation.length} together with the definition
\( \lambda _{T}=l^{y_{T}}=l^{1/\nu } \), we can calculate the correlation length
critical exponent as\begin{equation}
\label{thermal.exponent}
\nu =\frac{1}{y_{T}}=\frac{\ln l}{\ln \lambda _{T}}.
\end{equation}
The magnetic critical exponent is calculated from\begin{equation}
g_{p'}(K')\, H'=l^{d-y_{H}}\, g_{p}(K)\, H,
\end{equation}
 together with the scaling relation for the magnetic fields, \( H'=l^{y_{H}}\, H \),
as\begin{equation}
y_{H}=\frac{1}{2}\left( d+\frac{1}{\ln l}\ln \frac{g_{p}(K_{c})}{g_{p'}(K_{c})}\right) .
\end{equation}
 The only source of inaccuracy in these relations is in the finite sizes of
the clusters. However, it has been found that these kinds of approaches are
applicable to rather small systems and, sometimes, they are even more accurate
than other RG approaches, for clusters of adequate size\cite{PLASCAK1999}.
\begin{figure}
{\par\centering \subfigure[1-spin cluster.]{\includegraphics{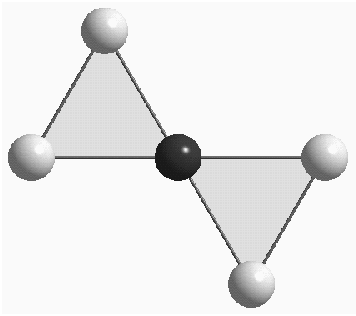}} \subfigure[3-spins cluster.]{\includegraphics{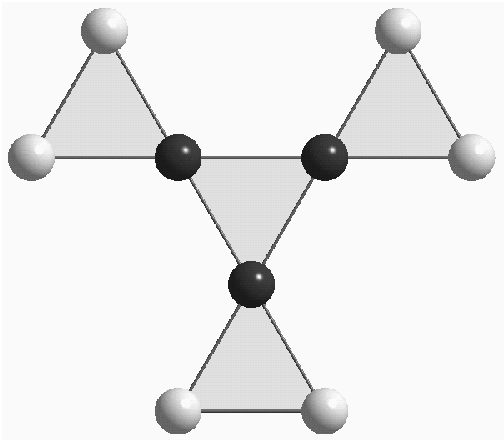}} \par}

\caption{\label{fig.kagome.clusters}Clusters used in this work for the \emph{kagomé}
lattice. Dark spheres represent spins belonging to the cluster itself and white
spheres the corresponding surrounding sites (those creating an effective field).}
\end{figure}

Even though the EFRG scheme has been successfully applied to a broad range of
problems, the original formulation of this method is not well suited to deal
with the effects of geometrical frustration, that is, magnetic systems with
a non--Bravais lattice, as those considered in this work. Therefore, our first
task, is to generalize this technique to include this class of systems. In order
to accomplish this goal, we must face two problems: on one hand, due to the
non--Bravais character of the lattices, we cannot characterize the full system
with only one order parameter; on the other hand, the finite clusters considered
have to have a minimum size such that all the order parameters variables are
included in the reduced trace, at least for one of the clusters, which involves
a more complicated mathematical problem. These difficulties, of course, are
not a particular flaw of the EFRG scheme, as they are present in all the real
space RG schemes, and are related to the fact that non-Bravais lattices are
non-invariant under a scale transformation. To cite an example, under a block
transformation, the \emph{kagomé} lattice does not transform in itself, but
in a triangular lattice.

The easiest way of solving these technical difficulties is by subdividing the
non-Bravais lattice in a number of Bravais lattices, and allowing for a different
order parameter in each sublattice. For example, in the case of the \emph{kagomé}
lattice, it can be considered as formed by \( n=3 \) interlocked triangular
lattices (\( \alpha =A,B,C \)), whereas in the pyrochlore case, it can considered
as formed by \( n=4 \) interlocked FCC lattices (\( \alpha =A,B,C,D \)).
\begin{figure}
{\par\centering \subfigure[1-spin cluster.]{\includegraphics{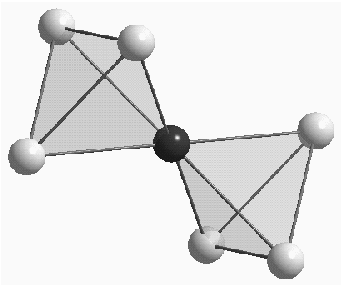}} \subfigure[4-spins cluster.]{\includegraphics{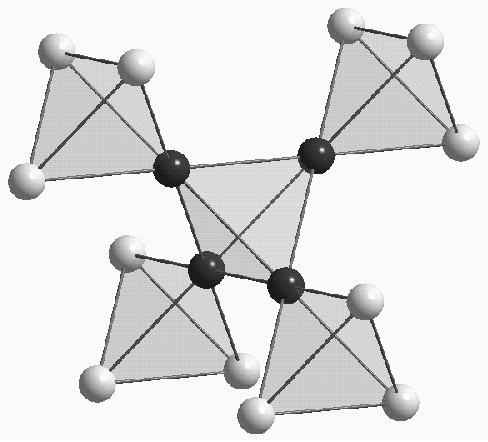}} \par}

\caption{\label{fig.pyrochlore.clusters}Clusters used in this work for the pyrochlore
lattice. Dark spheres represent spins belonging to the cluster itself and white
spheres the corresponding surrounding sites (those creating an effective field).}
\end{figure}

Next, we have to decide the size of the two finite clusters we are considering.
The most obvious choice is to consider a cluster formed by one single spin and
another one formed by \( p=3 \) spins for the \emph{kagomé} lattice (see Fig.~\ref{fig.kagome.clusters})
and \( p=4 \) spins for the pyrochlore lattice (see Fig.~\ref{fig.pyrochlore.clusters}).

The following step is to evaluate the expectation value of each of the order
parameters for these clusters by using the Callen--Suzuki identity, a task that
we carry out in the next section.

\section{Evaluation of the order parameters}

\subsection{1-spin cluster order parameter}

Let us consider\label{sec.order.parameters} the cluster defined by the ion
1 in sublattice \( \alpha  \). The Hamiltonian of this cluster is given by\begin{equation}
\mathcal{H}_{1\alpha }=K'\vec{s}_{1\alpha }\sum _{\beta \neq \alpha }\sum _{i}^{z}\vec{s}_{i\beta }=\vec{s}_{1\alpha }\cdot \vec{\xi }_{1\alpha },
\end{equation}
 where the subindex \( 1\alpha  \) makes reference to the fact that the Hamiltonian
corresponds to a cluster of 1 spin belonging to sublattice \( \alpha  \), and
\( \vec{\xi }_{1\alpha } \) stands for the symmetry breaking field acting on
the spin of the 1-spin cluster, which belong to sublattice \( \alpha  \). The
sum over the \( i \) index is performed over the NN of the \( 1\alpha  \)
spin, with \( z \) the number of NN in each of the sublattices (\( z=2 \)
for both the \emph{kagomé} and pyrochlore lattices). By using the Callen--Suzuki
identity we can calculate the order parameter in sublattice \( \alpha  \) for
a cluster formed by 1 spin\begin{equation}
\vec{m}_{1\alpha }=\left\langle \vec{s}_{1\alpha }\right\rangle =\left\langle \frac{Tr_{1\alpha }\vec{s}_{1\alpha }e^{\mathcal{H}_{1\alpha }}}{Tr_{1\alpha }e^{\mathcal{H}_{1\alpha }}}\right\rangle _{\mathcal{H}}.
\end{equation}
 As commented above, \( Tr_{1\alpha } \) represents the partial trace with
respect to the variables of the spin \( 1\alpha  \), and the symbol \( \left\langle \ldots \right\rangle _{\mathcal{H}} \)
stands for the usual canonical thermal average taken over the ensemble defined
by the complete Hamiltonian \( \mathcal{H} \). This equation can be put in
the more convenient form\begin{equation}
m_{1\alpha }^{\mu }=\left\langle \frac{\partial }{\partial \xi _{1\alpha }^{\mu }}\ln Tr_{1\alpha }e^{\mathcal{H}_{1\alpha }}\right\rangle _{\mathcal{H}},
\end{equation}
 where \( a^{\mu } \) stands for the \( \mu  \)-th component of the vector
\( \vec{a} \). A simple calculation yields\begin{equation}
\label{1spin.op}
\vec{m}_{1\alpha }=\widehat{n}_{\alpha }\left\langle \tanh \left( \widehat{n}_{\alpha }\cdot \vec{\xi }_{1\alpha }\right) \right\rangle _{\mathcal{H}}.
\end{equation}

Next, we make use of the operator identity\cite{HONMURA1979}\begin{equation}
\label{op.identity}
f(x+a)=e^{a\, D_{x}}f(x),
\end{equation}
 where \( D_{x}=\frac{\partial }{\partial x} \), to rewrite eq. \eqref{1spin.op}
in the more convenient form\begin{equation}
\vec{m}_{1\alpha }=\widehat{n}_{\alpha }\, \left\langle e^{\widehat{n}_{\alpha }\cdot \vec{\xi }_{1\alpha }D_{x}}\right\rangle _{\mathcal{H}}\left. \tanh x\right| _{x=0}.
\end{equation}

Finally, by making use of the van der Waerden identity\begin{equation}
e^{aS}=\cosh a+S\sinh a,\, \, \, \, \, \, S=\pm 1,
\end{equation}
 we can put\begin{equation}
e^{\widehat{n}_{\alpha }\cdot \vec{\xi }_{1\alpha }D_{x}}=\exp \left[ K'\widehat{n}_{\alpha }\cdot \sum _{\beta \neq \alpha }\sum _{i}^{z}\vec{s}_{i\beta }\, D_{x}\right] =\prod _{\beta \neq \alpha }\prod _{i}^{z}\left[ \rho _{x}+s_{i\beta }\nu _{x}\right] ,
\end{equation}
 where \( \rho _{x}=\cosh \left( \mathcal{K}'D_{x}\right)  \) and \( \nu _{x}=\sinh \left( \mathcal{K}'D_{x}\right)  \),
with \( \mathcal{K}'=K'\cos \theta =K'\widehat{n}_{\alpha }\cdot \widehat{n}_{\beta } \).
The expression of \( \mathcal{K}' \) takes into account the fact that, for
the structures considered in this work, the angle between the directions of
anisotropy is the same for any pair of neighboring spins.

Then, the order parameter for the 1-spin cluster can be put in the form
\begin{equation} 
\label{order1} \vec{m}_{1\alpha }=\widehat{n}_{\alpha }\left\langle \prod _{\beta \neq \alpha }\prod _{i}^{z}\left[ \rho _{x}+s_{i\beta }\nu _{x}\right] \right\rangle _{\mathcal{H}}\left. \tanh x\right| _{x=0}.
\end{equation}
For example, for the \emph{kagomé} lattice, the previous expression takes the
form \begin{multline}
\vec{m}_{1\alpha }=\widehat{n}_{\alpha }\left[ 2\rho _{x}^{3}\nu _{x}\left( \left\langle s_{1\beta }\right\rangle _{\mathcal{H}}+\left\langle s_{1\gamma }\right\rangle _{\mathcal{H}}\right) \right.\\ \left.+\rho _{x}^{2}\nu _{x}^{2}\left( \left\langle s_{1\beta }s_{2\beta }\right\rangle _{\mathcal{H}}+\left\langle s_{1\gamma }s_{2\gamma }\right\rangle _{\mathcal{H}}+2\left\langle s_{1\beta }s_{1\gamma }\right\rangle _{\mathcal{H}}+2\left\langle s_{1\beta }s_{2\gamma }\right\rangle _{\mathcal{H}}\right)\right.\\ \left.+2\rho _{x}\nu _{x}^{3}\left( \left\langle s_{1\beta }s_{1\gamma }s_{2\gamma }\right\rangle _{\mathcal{H}}+\left\langle s_{1\beta }s_{1\gamma }s_{2\beta }\right\rangle _{\mathcal{H}}\right) +\nu _{x}^{4}\left\langle s_{1\beta }s_{1\gamma }s_{2\beta }s_{2\gamma }\right\rangle _{\mathcal{H}}\right] \left. \tanh x\right| _{x=0},\\
(\alpha\neq\beta\neq\gamma).
\end{multline}These kind of equations, though exact, are intractable, and we
have to resort to some kind of approximation. The simplest one corresponds to
neglect 2-spin and higher order correlations\cite{HONMURA1979}, that is\begin{equation}
\label{decoupling}
\left\langle s_{i\alpha }s_{j\beta }\ldots s_{k\gamma }\right\rangle _{\mathcal{H}}\simeq \left\langle s_{i\alpha }\right\rangle _{\mathcal{H}}\left\langle s_{j\beta }\right\rangle _{\mathcal{H}}\ldots \left\langle s_{k\gamma }\right\rangle _{\mathcal{H}},
\end{equation}
 which is equivalent to neglect terms of the order of \( K_{c}^{2} \) and higher
in the calculation of the critical points. It must be noted, however, that this
approximation is superior to the standard MF theory, since it takes exactly
into account the relation \( \left\langle s_{i\beta }^{2}\right\rangle =1 \)
(\( i\neq 1 \)), through the van der Waerden identity.

Applying \eqref{decoupling} to \eqref{order1}, and to first order in \( \left\langle s_{i\alpha }\right\rangle _{\mathcal{H}}=b_{1\alpha } \)
(which constitutes a good approximation near the critical point), where the
subindex 1 means that these symmetry breaking fields have been obtained from
the 1-spin cluster, we arrive to the final form of the order parameter in sublattice
\( \alpha  \)\begin{equation}
\vec{m}_{1\alpha }=\widehat{n}_{\alpha }\, A(\mathcal{K}')\sum _{\beta \neq \alpha }b_{1\beta },
\end{equation}
 where\begin{equation}
\label{aes}
A(\mathcal{K}')=z\, \rho _{x}^{z(p-1)-1}\nu _{x}\left. \tanh x\right| _{x=0},
\end{equation}
 which, after a straightforward calculation it is found to be\begin{equation}
\label{eq1}
A(\mathcal{K}')=\frac{2\tanh (2\mathcal{K}')+\tanh (4\mathcal{K}')}{4},
\end{equation}
 for the \emph{kagomé} lattice (\( z=2 \) and \( p=3 \)), and\begin{equation}
\label{eq2}
A(\mathcal{K}')=\frac{5\tanh (2\mathcal{K}')+4\tanh (4\mathcal{K}')+\tanh (6\mathcal{K}')}{16},
\end{equation}
 for the pyrochlore lattice (\( z=2 \) and \( p=4 \)).

\subsection{\protect\protect\( p\protect \protect \)-spin cluster order parameter}

Let us consider now a cluster formed by \( p \) spins, each of one belonging
to a different sublattice (therefore, we will have \( p=3 \) for the \emph{kagomé}
lattice and \( p=4 \) for the pyrochlore lattice). The Hamiltonian of such
a cluster can be put as\begin{equation}
\mathcal{H}_{p}=K\sum _{\alpha \neq \beta }\vec{s}_{1\alpha }\cdot \vec{s}_{1\beta }+\sum _{\alpha }\vec{s}_{1\alpha }\cdot \vec{\xi }_{p\alpha }.
\end{equation}
In this Hamiltonian, the first term represents the interaction between the spins
inside the cluster of size \( p \), whereas the second term represents the
interaction of those same spins with the symmetry breaking fields\begin{equation}
\vec{\xi }_{p\alpha }=K\sum _{\beta \neq \alpha }\sum _{i}^{z-1}\vec{s}_{i\beta },
\end{equation}
 created by the rest of spins outside the cluster. The sum over \( i \) goes
over all the NN spins except the ones already included in the cluster. The order
parameter for each of the sublattices included in the cluster, obtained from
the \( p \)-spin cluster can be now put as\begin{equation}
\vec{m}_{p\alpha }=\left\langle \frac{Tr_{1A,1B,1C\ldots }\vec{s}_{1\alpha }e^{\mathcal{H}_{p}}}{Tr_{1A,1B,1C\ldots }e^{\mathcal{H}_{p}}}\right\rangle _{\mathcal{H}},
\end{equation}
 or, more conveniently\begin{equation}
\label{order2}
m_{p\alpha }^{\mu }=\left\langle \frac{\partial }{\partial \xi _{p\alpha }^{\mu }}\ln Tr_{1A,1B,1C\ldots }e^{\mathcal{H}_{p}}\right\rangle _{\mathcal{H}}.
\end{equation}
 In these expressions, \( Tr_{1A,1B,1C\ldots } \) represents the trace over
the spin degrees of freedom in the \( p \)-spin cluster.

Even though the discussion can be continued for a cluster with an arbitrary
number of spins, the mathematical details get very cumbersome. Therefore, we
will carry on separately the calculation of the order parameter for the \emph{kagomé}
and the pyrochlore lattices.

\subsubsection{Kagomé \emph{lattice}}

In the case of the \emph{kagomé} lattice, we will consider a triangular cluster
defined by \( p=3 \) spins. In doing so, we can easily calculate the reduced
partition function appearing in \eqref{order2} \begin{multline}
Tr_{1A,1B,1C}e^{\mathcal{H}_{p}}=Z_{3}(\mathcal{K},\Pi _{A},\Pi _{B},\Pi _{C})=  \cosh (\Pi _{A}+\Pi _{B}+\Pi _{C})\\
 +e^{-4\mathcal{K}}\left[ \cosh (\Pi _{A}-\Pi _{B}+\Pi _{C})+\cosh (\Pi _{A}+\Pi _{B}-\Pi _{C})+\cosh (\Pi _{A}-\Pi _{B}-\Pi _{C})\right],
\end{multline} where \( \Pi _{\alpha }=\widehat{n}_{\alpha }\cdot \vec{\xi }_{3\alpha } \)
and \( \mathcal{K}=K\cosh \theta  \). Taking into account the relation\begin{equation}
\frac{\partial }{\partial \xi _{3\alpha }^{\mu }}\Pi _{\alpha }=n_{\alpha }^{\mu },
\end{equation}
 we obtain the expression of the order parameter for the 3-spin cluster \begin{multline}
\label{order.parameter.kagome}
\vec{m}_{3\alpha }=\left\langle \frac{\sinh (\Pi _{A}+\Pi _{B}+\Pi _{C})+e^{-4\mathcal{K}}\left[ \pm \sinh (\Pi _{A}-\Pi _{B}+\Pi _{C})\right.}{Z_{3}(\mathcal{K},\Pi _{A},\Pi _{B},\Pi _{C})}\right.\\
+\left.\frac{\left.\pm \sinh (\Pi _{A}+\Pi _{B}-\Pi _{C})\pm \sinh (\Pi _{A}-\Pi _{B}-\Pi _{C})\right] }{Z_{3}(\mathcal{K},\Pi _{A},\Pi _{B},\Pi _{C})}\right\rangle _{\mathcal{H}}\widehat{n}_{\alpha },
\end{multline} where the \( + \) or \( - \) sign corresponds to the sign of
\( \Pi _{\alpha } \) in the argument of the corresponding \( \sinh  \). To
quote an example, the order parameter in sublattice \( B \) is given by \begin{multline}
\vec{m}_{3B }=\left\langle \frac{\sinh (\Pi _{A}+\Pi _{B}+\Pi _{C})+e^{-4\mathcal{K}}\left[ - \sinh (\Pi _{A}-\Pi _{B}+\Pi _{C})\right.}{Z_{3}(\mathcal{K},\Pi _{A},\Pi _{B},\Pi _{C})}\right.\\
+\left.\frac{\left.+ \sinh (\Pi _{A}+\Pi _{B}-\Pi _{C})- \sinh (\Pi _{A}-\Pi _{B}-\Pi _{C})\right] }{Z_{3}(\mathcal{K},\Pi _{A},\Pi _{B},\Pi _{C})}\right\rangle _{\mathcal{H}}\widehat{n}_{B },
\end{multline}

By making use again of the operator identity \eqref{op.identity}, we can express
the order parameters as\begin{equation}
\label{order3}
\vec{m}_{3\alpha }=\widehat{n}_{\alpha }\left\langle e^{\Pi _{A}D_{x}}e^{\Pi _{B}D_{y}}e^{\Pi _{C}D_{z}}\right\rangle _{\mathcal{H}}\left. g_{\alpha }(x,y,z)\right| _{x=y=z=0},
\end{equation}
 where \( g_{\alpha }(x,y,z) \) is obtained by making the substitutions \( \Pi _{A}\to x \),
\( \Pi _{B}\to y \), and \( \Pi _{C}\to z \) in \eqref{order.parameter.kagome}.
For example \begin{multline}
\label{ga.kagome}
g_{B}(x,y,z)=\\
\frac{\sinh (x+y+z)+e^{-4\mathcal{K}}\left[ -\sinh (x-y+z)+\sinh (x+y-z)-\sinh (x-y-z)\right] }{\cosh (x+y+z)+e^{-4\mathcal{K}}\left[ \cosh (x-y+z)+\cosh (x+y-z)+\cosh (x-y-z)\right] }.
\end{multline}

\begin{figure}
{\par\centering \resizebox*{3in}{!}{\includegraphics{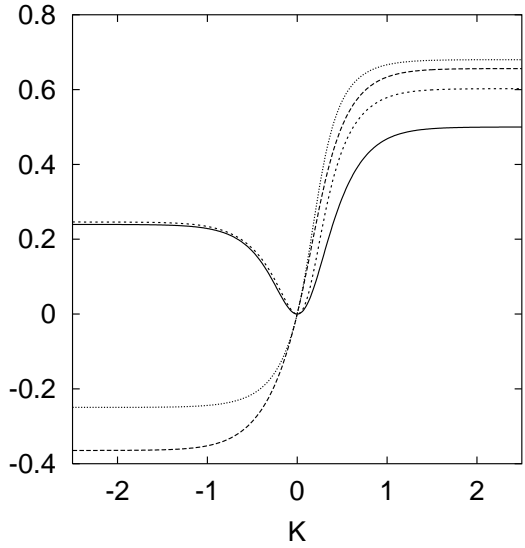}} \par}

\caption{\label{fig.bes}\protect\protect\( \mathcal{K}\protect \protect \) dependence
of \protect\protect\( B_{1}(\mathcal{K})\protect \protect \) (solid line),
\protect\protect\( B_{2}(\mathcal{K})\protect \protect \) (long dashed line), \protect\protect\( C_{1}(\mathcal{K})\protect \protect \)
(short dashed line), and \protect\protect\( C_{2}(\mathcal{K})\protect \protect \)
(dotted line), given by \eqref{b1}, \eqref{b2}, \eqref{eq5}, and \eqref{eq6},
respectively.}
\end{figure}

By expanding the exponentials of sums in \eqref{order3} as products of exponentials,
and making use again of the van der Waerden identity and the decoupling of the
correlations, we arrive, to first order in the symmetry breaking fields \( b_{3\alpha }=\left\langle s_{i\alpha }\right\rangle _{\mathcal{H}} \),
to the following system of equations\begin{equation}
\label{kagome.magnetization.zero.field}
\vec{m}_{3\alpha }=\left[ B_{1}(\mathcal{K})\, b_{3\alpha }+B_{2}(\mathcal{K})(b_{3\beta }+b_{3\gamma })\right] \widehat{n}_{\alpha },\, \, \, \, (\alpha \neq \beta \neq \gamma ),
\end{equation}
 where\begin{equation}
\label{b1}
B_{1}(\mathcal{K})=\rho _{x}^{2}\rho _{y}\rho _{z}(\nu _{y}\rho _{z}+\rho _{y}\nu _{z})\left. g_{A}(x,y,z)\right| _{x=y=z=0},
\end{equation}
 and\begin{equation}
\label{b2}
B_{2}(\mathcal{K})=\rho _{y}^{2}\rho _{x}\rho _{z}(\nu _{x}\rho _{z}+\rho _{x}\nu _{z})\left. g_{A}(x,y,z)\right| _{x=y=z=0},
\end{equation}
 with \( \rho _{\mu }=\cosh (\mathcal{K}D_{\mu }) \) and \( \nu _{\mu }=\sinh (\mathcal{K}D_{\mu }) \),
\( \mu =x,y,z \). After applying the differential operators in the previous
expressions, we arrive to the final form for these coefficients \begin{multline}
\label{eq3}
B_{1}(\mathcal{K})=\frac{1}{12}\left[ \left( \frac{1}{2\cosh (4\mathcal{K})+3e^{-4\mathcal{K}}-1}+\frac{3}{\cosh (4\mathcal{K})+e^{-4\mathcal{K}}(2+\cosh (4\mathcal{K}))}\right.\right.\\
\left.\left.-\frac{6e^{-4\mathcal{K}}}{1+e^{-4\mathcal{K}}(1+2\cosh (4\mathcal{K}))}\right) \sinh (4\mathcal{K})+\frac{4\tanh (2\mathcal{K})}{1+3e^{-4\mathcal{K}}},\right] 
\end{multline} \begin{multline}
\label{eq4}
B_{2}(\mathcal{K})=\frac{1}{24}\left[ \left( \frac{2}{2\cosh (4\mathcal{K})+3e^{-4\mathcal{K}}-1}+\frac{3(2+e^{-4\mathcal{K}})}{\cosh (4\mathcal{K})+e^{-4\mathcal{K}}(2+\cosh (4\mathcal{K}))}\right.\right.\\
\left.\left.+\frac{3e^{-4\mathcal{K}}}{1+e^{-4\mathcal{K}}(1+2\cosh (4\mathcal{K}))}\right) \sinh (4\mathcal{K})+\frac{4(2+3e^{-4\mathcal{K}})\tanh (2\mathcal{K})}{1+3e^{-4\mathcal{K}}},\right] 
\end{multline} Certainly, it is not obvious to see the behavior of \( B_{1} \)
and \( B_{2} \) from the above expressions, for they are depicted in Fig.~\ref{fig.bes}. 

\subsubsection{Pyrochlore \emph{lattice}.}

The only difference with the previous case is that the cluster contains \( p=4 \)
spins, corresponding to 4 different sublattices (\( A,B,C \), and \( D \)).
Therefore, we will not present the details of the calculation here, but only
the main results.

The reduced partition function is given in this case by \begin{multline}
Z_{4}(\mathcal{K},\Pi _{A},\Pi _{B},\Pi _{C},\Pi _{D})=\cosh (\Pi _{A}+\Pi _{B}+\Pi _{C}+\Pi _{D})\\
+e^{-6\mathcal{K}}\left[ \cosh (\Pi _{A}-\Pi _{B}+\Pi _{C}+\Pi _{D})+\cosh (\Pi _{A}+\Pi _{B}-\Pi _{C}+\Pi _{D})\right.\\
\left.+\cosh (\Pi _{A}+\Pi _{B}+\Pi _{C}-\Pi _{D})+\cosh (\Pi _{A}-\Pi _{B}-\Pi _{C}-\Pi _{D})\right]\\
+e^{-8\mathcal{K}}\left[ \cosh (\Pi _{A}-\Pi _{B}-\Pi _{C}+\Pi _{D})+\cosh (\Pi _{A}-\Pi _{B}+\Pi _{C}-\Pi _{D})\right.\\
\left.+\cosh (\Pi _{A}+\Pi _{B}-\Pi _{C}-\Pi _{D})\right],
\end{multline} where the definition of \( \Pi _{\alpha } \) is as given above.

The expression of the order parameter for this cluster is given by \begin{multline}
\label{order.full}
\vec{m}_{4\alpha }=\left[ \frac{\sinh (\Pi _{A}+\Pi _{B}+\Pi _{C}+\Pi _{D})}{Z_{4}(\mathcal{K},\Pi _{A},\Pi _{B},\Pi _{C},\Pi _{D})}\right.\\
+e^{-6\mathcal{K}}\left(\frac{\pm \sinh (\Pi _{A}-\Pi _{B}+\Pi _{C}+\Pi _{D})\pm \sinh (\Pi _{A}+\Pi _{B}-\Pi _{C}+\Pi _{D})}{Z_{4}(\mathcal{K},\Pi _{A},\Pi _{B},\Pi _{C},\Pi _{D})}\right.\\
\left.+\frac{\pm \sinh (\Pi _{A}+\Pi _{B}+\Pi _{C}-\Pi _{D})\pm \sinh (\Pi _{A}-\Pi _{B}-\Pi _{C}-\Pi _{D})}{Z_{4}(\mathcal{K},\Pi _{A},\Pi _{B},\Pi _{C},\Pi _{D})}\right)\\
+e^{-8\mathcal{K}}\left(\frac{\pm \sinh (\Pi _{A}-\Pi _{B}-\Pi _{C}+\Pi _{D})\pm \sinh (\Pi _{A}-\Pi _{B}+\Pi _{C}-\Pi _{D})}{Z_{4}(\mathcal{K},\Pi _{A},\Pi _{B},\Pi _{C},\Pi _{D})}\right.\\
\left.\left.+\frac{\pm \sinh (\Pi _{A}+\Pi _{B}-\Pi _{C}-\Pi _{D})}{Z_{4}(\mathcal{K},\Pi _{A},\Pi _{B},\Pi _{C},\Pi _{D})}\right)\right]\widehat{n}_{\alpha }, 
\end{multline} where, as before, the \( + \) or \( - \) sign corresponds to
the sign of the corresponding \( \Pi _{\alpha } \) in the argument of the \( \sinh  \).

By making use of the operator identity \eqref{op.identity}, the van der Waerden
identity, and the decoupling approximation, we arrive to the expression of the
4-spin cluster order parameter up to first order in the symmetry breaking fields
\( b_{4\alpha }=\left\langle s_{i\alpha }\right\rangle _{\mathcal{H}} \), which
is given by the system of equations\begin{equation}
\label{order4}
\vec{m}_{4\alpha }=\left[ C_{1}(\mathcal{K})\, b_{4\alpha }+C_{2}(\mathcal{K})\left( b_{4\beta }+b_{4\gamma }+b_{4\delta }\right) \right] \widehat{n}_{\alpha },\, \, \, \, (\alpha \neq \beta \neq \gamma \neq \delta ),
\end{equation}
 where\begin{equation}
\label{eq5}
C_{1}(\mathcal{K})=\rho _{x}^{3}\rho _{y}^{2}\rho _{z}^{2}\rho _{u}^{2}(\nu _{y}\rho _{z}\rho _{u}+\rho _{y}\nu _{z}\rho _{u}+\rho _{y}\rho _{z}\nu _{u})\left. g_{A}(x,y,z,u)\right| _{x=y=z=u=0},
\end{equation}
 and\begin{equation}
\label{eq6}
C_{2}(\mathcal{K})=\rho _{x}^{2}\rho _{y}^{3}\rho _{z}^{2}\rho _{u}^{2}(\nu _{x}\rho _{z}\rho _{u}+\rho _{x}\nu _{z}\rho _{u}+\rho _{x}\rho _{z}\nu _{u})\left. g_{A}(x,y,z,u)\right| _{x=y=z=u=0}.
\end{equation}
 The function \( g_{\alpha }(x,y,z,u) \) is obtained from the coefficient of
\( \widehat{n}_{\alpha } \) in \eqref{order.full} with the substitutions \( \Pi _{A}\rightarrow x \),
\( \Pi _{B}\rightarrow y \), \( \Pi _{C}\rightarrow z \), and \( \Pi _{D}\rightarrow u \).

The evaluation of \( C_{1} \) and \( C_{2} \) is straightforward, though rather
lengthy, and the final expressions are of little use, due to their length. For
that reason, we have preferred not to quote them here, but illustrate their
behavior by Fig.~\ref{fig.bes}.

\section{Scaling relations and Critical behavior}

Let us \label{sec.critical.behavior}summarize the results obtained so far.
The order parameters of each sublattice for the 1-spin cluster are related by
the system of equations\begin{equation}
\label{system1}
\vec{m}_{1\alpha }=A(\mathcal{K}')\sum _{\beta \neq \alpha }b_{1\beta }\, \widehat{n}_{\alpha },
\end{equation}
 whereas the order parameters of each sublattice for the \( p \)-spin cluster
are related by the equations\begin{equation}
\label{system2}
\vec{m}_{p\alpha }=\left[ \Phi (\mathcal{K})\, b_{p\alpha }+\Theta (\mathcal{K})\sum _{\beta \neq \alpha }b_{p\beta }\right] \widehat{n}_{\alpha },
\end{equation}
 where \( A(\mathcal{K}') \), \( \Phi (\mathcal{K}) \), and \( \Theta (\mathcal{K}) \),
are given by expressions \eqref{eq1}, \eqref{eq3}, \eqref{eq4}, and \eqref{eq2},
\eqref{eq5}, \eqref{eq6}, for the \emph{kagomé} and pyrochlore lattices, respectively.

By making use of the scaling hypothesis \eqref{finite.scaling2} for both the
order parameter and the symmetry breaking fields \cite{PLASCAK1999} we arrive
to a system of equations for the symmetry breaking fields

\begin{equation}
\label{final.system}
\Phi (\mathcal{K})\, b_{p\alpha }+\left[ \Theta (\mathcal{K})-A(\mathcal{K}')\right] \sum _{\beta \neq \alpha }b_{p\beta }=0,
\end{equation}
 where \( \Phi  \), \( \Theta  \), and \( A \) are to be replaced by the
corresponding expressions, and \( \alpha ,\beta =A,B, \) and \( C \) for the
\emph{kagomé} lattice and \( \alpha ,\beta =A,B,C, \) and \( D \) for the
pyrochlore lattice. This homogeneous system has a non trivial solution if its
characteristic determinant is zero, which gives the condition\begin{equation}
\left[ (p-1)A(\mathcal{K}')-\Phi (\mathcal{K})-(p-1)\Theta (\mathcal{K})\right] \left[ A(\mathcal{K}')+\Phi (\mathcal{K})-\Theta (\mathcal{K})\right] ^{p-1}=0,
\end{equation}
 which has two solutions that give us the corresponding RG recursion relations
for the coupling constants.

\begin{figure}
{\par\centering \subfigure[\emph{Kagom\'{e}} lattice.]{\resizebox*{8cm}{!}{\includegraphics{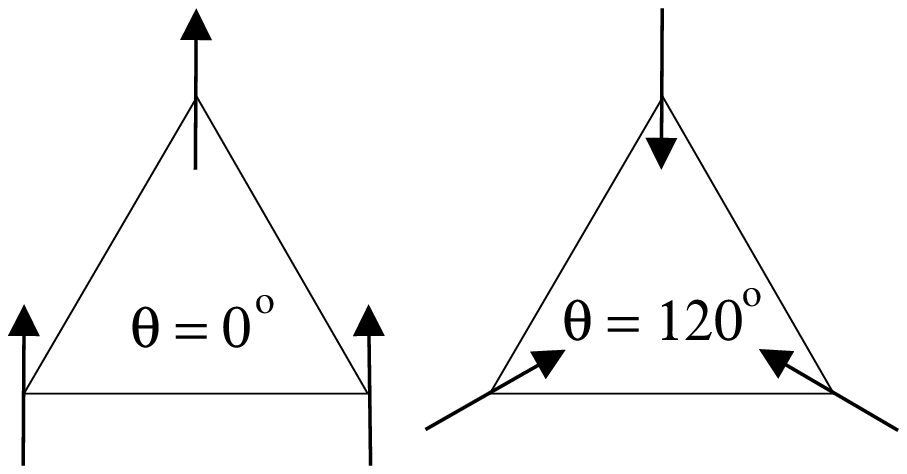}}} \subfigure[Pyrochlore lattice.]{\resizebox*{8cm}{!}{\includegraphics{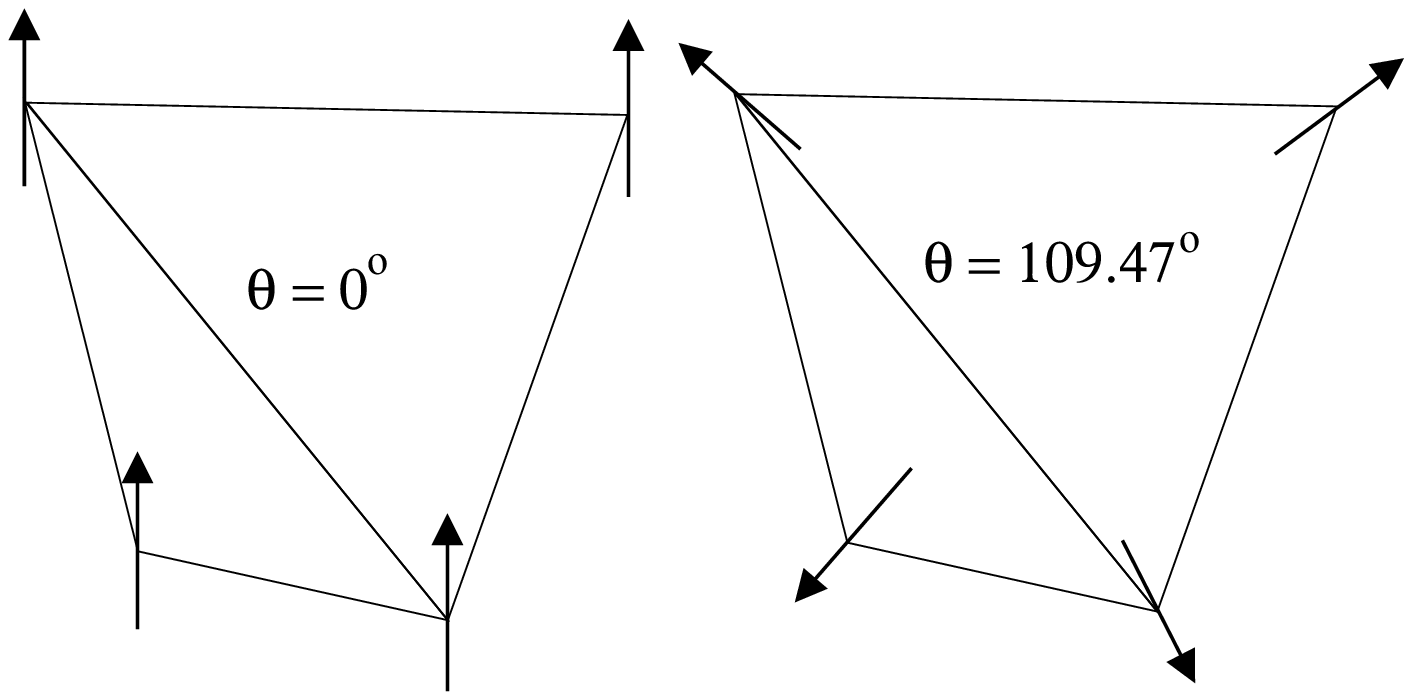}}} \par}

\caption{\label{fig.pointing.inside.outside}Type A configuration for the \emph{kagomé}
and pyrochlore lattices at \protect\( T=0\protect \). The \protect\protect\( \theta =0\protect \protect \)
case corresponds to the standard Ising model, whereas \protect\protect\( \theta \neq 0\protect \protect \)
corresponds to anisotropic Ising interactions (see text).}
\end{figure}

The first one corresponds to\begin{equation}
\label{cond1}
\Phi (\mathcal{K})+(p-1)\Theta (\mathcal{K})=(p-1)A(K').
\end{equation}
 In this case, the solution of the system of equations is given by (notice that
we have omitted the index corresponding to the size of the cluster)\begin{equation}
b_{A}=b_{B}=b_{C}=\ldots 
\end{equation}
 or, in terms of the modulus of the order parameter,\begin{equation}
\label{typeA}
m_{A}=m_{B}=m_{C}=\ldots ,
\end{equation}
 and corresponds to a configuration in which all the spins point {}``out{}''
or all the spins point {}``in{}'', as has been depicted in Fig.~\ref{fig.pointing.inside.outside}.
We will term this configuration {}``Type A{}''.

The second solution occurs if\begin{equation}
\label{cond2}
\Theta (\mathcal{K})-\Phi (\mathcal{K})=A(\mathcal{K}'),
\end{equation}
 which leads to a configuration in which\begin{equation}
\sum _{\alpha }b_{\alpha }=0,
\end{equation}
 or, in terms of the order parameters\begin{equation}
\label{typeB}
\sum _{\alpha }m_{\alpha }=0.
\end{equation}
 We will term this configuration as {}``Type B{}''.

It is important to notice that through the introduction of the effective couplings
\( \mathcal{K}' \) and \( \mathcal{K} \), we can consider both the standard
Ising Hamiltonian in these geometries and the corresponding spin ice problem
on equal footing or, equivalently, we can map the spin ice problem in these
geometries to the standard Ising model. Indeed, the standard Ising model corresponds
to taking \( \cos \theta =\widehat{n}_{\alpha }\cdot \widehat{n}_{\beta }=1 \),
that is, the spins point along the same direction, whereas the case \( \theta =120^{\circ } \)
for the \emph{kagomé} lattice corresponds to the 2D spin ice analog commented
above. The case \( \theta =109.47^{\circ } \) is the spin ice problem in the
pyrochlore lattice. 

\begin{figure}
{\par\centering \subfigure[\emph{Kagom\'{e}} lattice.]{\resizebox*{8cm}{!}{\includegraphics{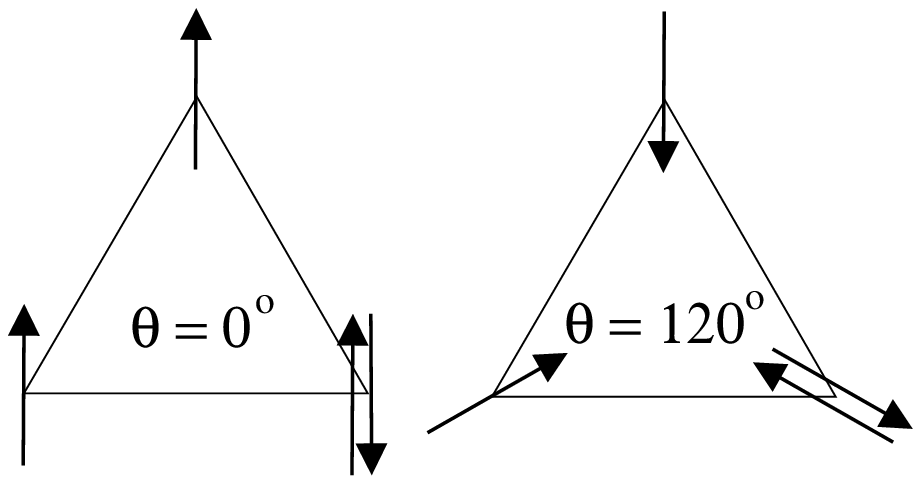}}} \subfigure[Pyrochlore lattice.]{\resizebox*{8cm}{!}{\includegraphics{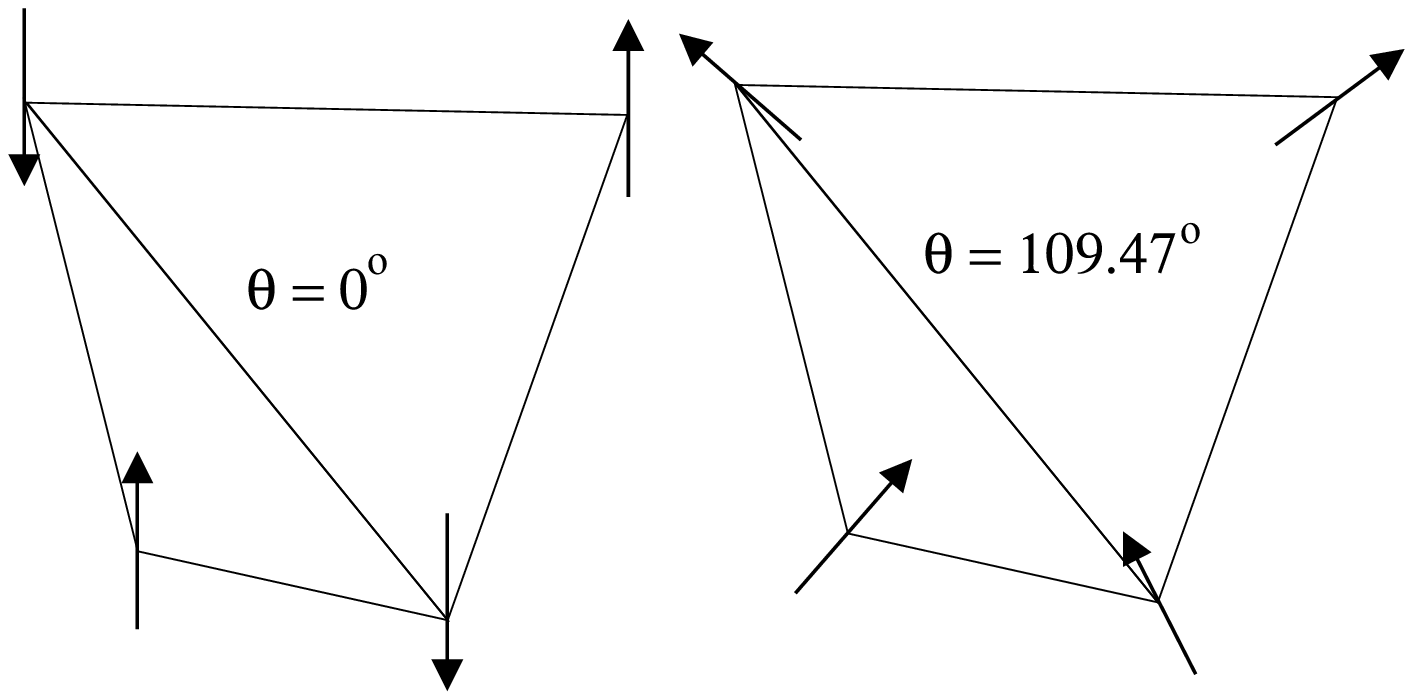}}} \par}

\caption{\label{fig.spin.ice.conf}Type B configuration for the \emph{kagomé} and pyrochlore
lattices at \protect\( T=0\protect \). The \protect\protect\( \theta =0\protect \protect \)
case corresponds to the standard antiferromagnetic Ising model, whereas \protect\protect\( \theta \neq 0\protect \protect \)
corresponds to anisotropic Ising interactions (see text).}
\end{figure}

In the case \( \theta =0^{\circ } \), the Type A configuration corresponds
to ferromagnetic order, whereas the Type B configuration is a generalization
of antiferromagnetic order. Moreover, in the pyrochlore case, and for Ising
spins pointing along \( \left\langle 1,1,1\right\rangle  \)-type directions,
the last configuration contains, among others more general, the ground state
configuration of the spin ice system, in which two spins point {}``in{}''
and two spins point {}``out{}'' (see Fig.~\ref{fig.spin.ice.conf}), and
reflects the degeneracy of the ground state due to the geometrical frustration.
It is important to notice that it impossible to satisfy condition \eqref{typeB}
in the \emph{kagomé} case at \( T=0 \), which has been noted in Fig.~\ref{fig.spin.ice.conf}
by a double spin in one of the corners.

\begin{figure}
{\par\centering \subfigure[Type A configuration.]{\resizebox*{3in}{!}{\includegraphics{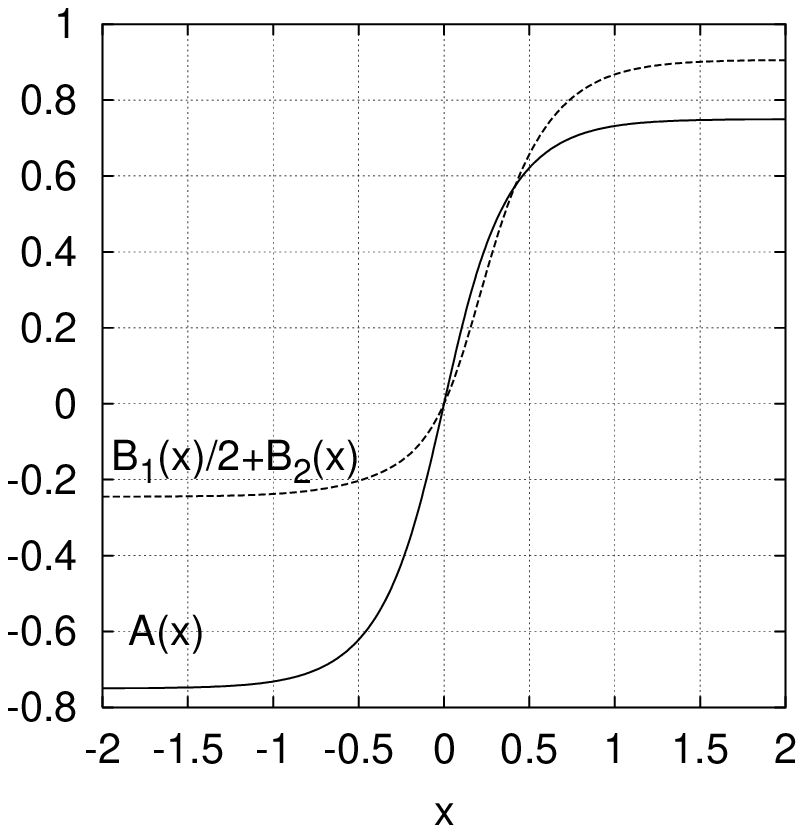}}} \subfigure[Type B configuration.]{\resizebox*{3in}{!}{\includegraphics{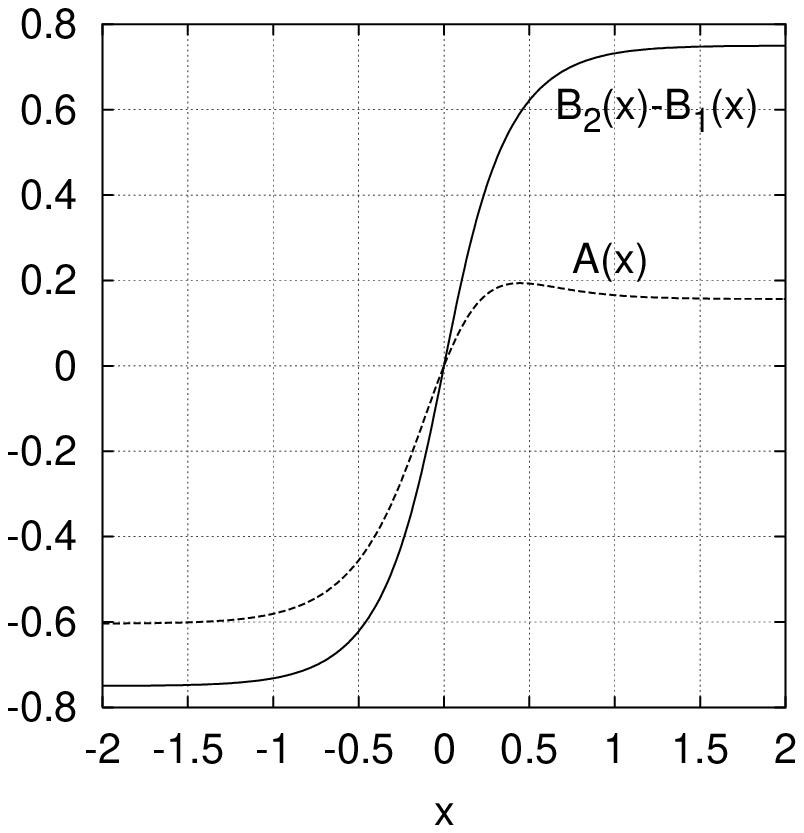}}} \par}

\caption{\label{fig.cond1}Graphical estimation of the fixed points of the RG transformation
for the \emph{kagomé} lattice. \protect\( A(x)\protect \) is given by \eqref{eq1},
whereas \protect\( B_{1}(x)\protect \) and \protect\( B_{2}(x)\protect \)
are given by \eqref{b1} and \eqref{b2}, respectively.}
\end{figure}

\begin{figure}
{\par\centering \subfigure[Type A configuration.]{\resizebox*{3in}{!}{\includegraphics{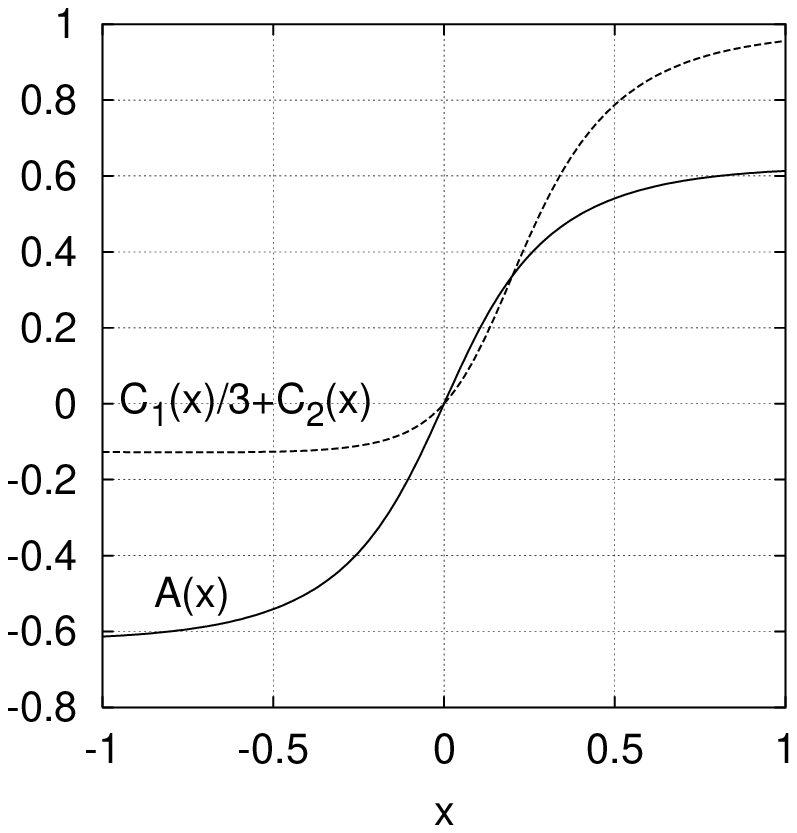}}} \subfigure[Type B configuration.]{\resizebox*{3in}{!}{\includegraphics{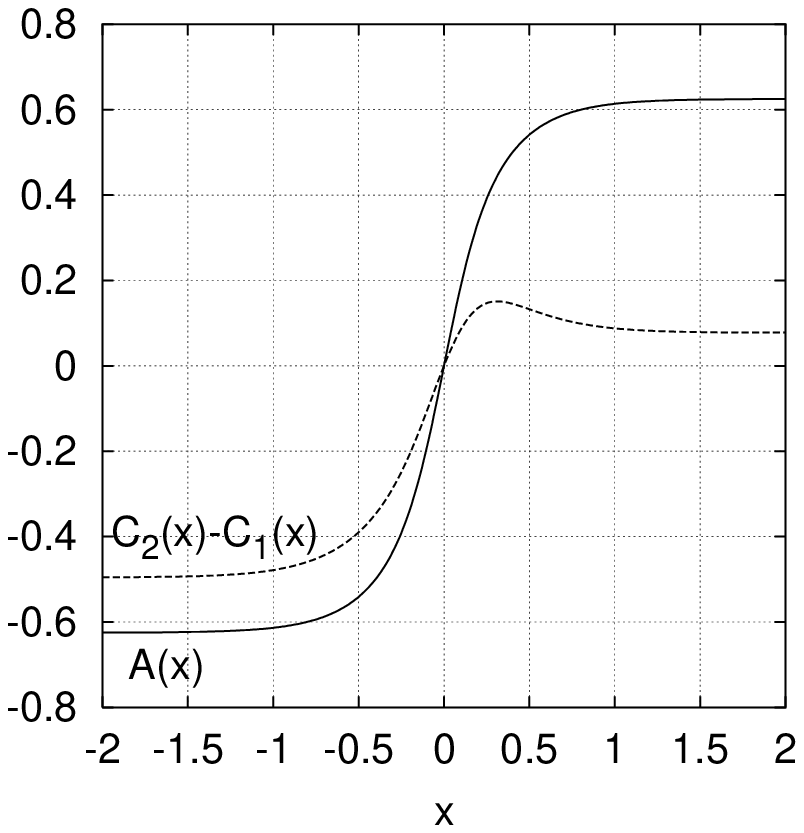}}} \par}

\caption{\label{fig.cond2}Graphical estimation of the fixed points of the RG transformation
for the pyrochlore lattice. \protect\( A(x)\protect \) is given by \eqref{eq2},
whereas \protect\( C_{1}(x)\protect \) and \protect\( C_{2}(x)\protect \)
are given by \eqref{eq5} and \eqref{eq6}, respectively.}
\end{figure}

Of course, apart from obtaining the possible spin configurations, it is even
more important to study if they are thermally stable, that is, if they really
occur. To this end, we have to see if there is any value \( \mathcal{K}'=\mathcal{K}=\mathcal{K}_{c} \)
that satisfies conditions \eqref{cond1} and \eqref{cond2}, that is, if the RG
recursion relations defined by \eqref{cond1} and \eqref{cond2} have any non-trivial
fixed points. The simplest way of solving these equations is by using graphical
methods, as has been done in Figs.~\ref{fig.cond1} and \ref{fig.cond2} for
the \emph{kagomé} and pyrochlore lattices, respectively. 

From these figures, we can see that for the Type B configuration, the only fixed
point of the RG transformation is the trivial one, corresponding to a paramagnetic
configuration (\( \mathcal{K}=\mathcal{K}'=0 \)). However, there is a non trivial
fixed point for the Type A configuration (\( \mathcal{K}_{c}=0.415 \) for the
\emph{kagomé} lattice and \( \mathcal{K}_{c}=0.23 \) for the pyrochlore lattice),
depending on the sign of the effective coupling or, in other words, depending
on whether the effective coupling is ferro- or antiferromagnetic. These results
have been collected in Table~\ref{table1}, for both the \emph{kagomé} and
pyrochlore lattices, and for both the standard Ising model and the spin ice
problem. Also, the exact results known for the \emph{kagomé} lattice\cite{SYOZI1972}
and the ones obtained from MC data for the pyrochlore lattice\cite{BRAMWELL1998}
are quoted for comparison. It can be seen that the estimates obtained from this
RG scheme are in very good agreement with the exact values, in spite of the
crudeness of the decoupling approximation and the fact that we have considered
clusters of small size. This reflects the well known fact that, in this GFAF
systems, correlations extending outside the frustrated unit are relatively small\cite{Canals98,Canals00a}.
On the other hand, correlations inside the unit are considered to some extent
by the present technique. In any case, the quality of the estimates can be systematically
improved by considering larger units as, in this way, we include further correlations.
However, the improvement in the estimate of the critical point is not commensurate
with the additional algebraic difficulty involved. Another important point is
that the EFRG method distinguishes among the various different geometries for
a given number of NN, in contrast with standard MF approaches, which do not
distinguish, for example between a cubic lattice and a pyrochlore one\footnote{%
This is the case even if we consider the smallest possible clusters. See, for
example, I. P. Fittipaldi and D. F. de Alburquerque, J. Magn. Magn. Mat. \textbf{104-107},
236 (1992), where they studied the Ising model with bond dilution in both the
simple square and \emph{kagomé} lattices in the EFRG framework by using clusters
of 1 and 2 spins, respectively.
}.
\begin{table}
{\centering \begin{tabular}{ccc}
\hline 
\multicolumn{1}{c}{\( \mathcal{K}_{c} \)}&
\emph{Kagomé}&
Pyrochlore\\
\hline 
\hline 
\multicolumn{1}{l}{EFRG}&
\( 0.415 \)&
\( 0.230 \)\\
\multicolumn{1}{l}{GCC}&
\( 0.402 \)&
\( 0.227 \)\\
\multicolumn{1}{l}{Exact}&
\( 0.4643 \)\footnote{Ref.~\onlinecite{SYOZI1972}}&
\( 0.25 \)\footnote{Ref.~\onlinecite{BRAMWELL1998}}\\
\hline 
\end{tabular}\par}

\caption{\label{table1}Non-trivial fixed points of the RG transformation, and corresponding
critical points in the GCC approach (see text).}
\end{table}

\begin{table}
\begin{tabular}{c|c|cc|cc}
\hline 
&
&
\multicolumn{2}{c|}{ \emph{Kagomé}}&
\multicolumn{2}{c}{Pyrochlore }\\
\hline 
&
&
 \( K>0 \)&
 \( K<0 \)&
 \( K>0 \)&
 \( K<0 \)\\
\multicolumn{1}{l|}{EFRG}&
 \( \theta =0 \)&
 \( 0.415 \)&
 ---&
\( 0.230 \)&
 ---\\
 &
\( \theta =120^{\circ }(109.47^{\circ }) \)&
 ---&
 \( -0.830 \)&
 ---&
\( -0.690 \)\\
\hline 
\multicolumn{1}{l|}{GCC}&
\( \theta =0 \)&
\( 0.402 \)&
---&
\( 0.227 \)&
---\\
&
\( \theta =120^{\circ }(109.47^{\circ }) \)&
---&
\( -0.804 \)&
---&
\( -0.671 \)\\
\hline 
\end{tabular}

\caption{\label{table2}Critical points obtained from the EFRG and GCC methods for the
Ising, \emph{kagomé} spin ice, and spin ice pyrochlore. The long dashes indicate
the absence of a critical point.}
\end{table}

Regarding the question of geometrical frustration, which leads to the absence
of a transition to a long range ordered state at any finite temperature, both
the \emph{kagomé} and pyrochlore ferromagnetic Ising lattices are non-frustrated,
whereas the corresponding antiferromagnetic ones are frustrated, and remain
paramagnetic down to 0 K. In the spin ice case (\( \theta =120^{\circ } \)
for the \emph{kagomé} lattice and \( \theta =109.47^{\circ } \) for the pyrochlore),
however, the ferromagnetic case is frustrated, and does not order at any finite
temperature, whereas the antiferromagnetic one experiments a transition to a
long range ordered state at a finite temperature (see Table~\ref{table2}).
In any case, the ground state given by the {}``ice rule{}'' \cite{BRAMWELL1998,RAMIREZ2000}
never occurs in these systems. Furthermore, we can see now that the reason why
the ferromagnetic spin ice system is frustrated is that it can be mapped to
the standard Ising Hamiltonian with antiferromagnetic effective interaction,
\( \mathcal{K}<0 \), which is known to be frustrated.

\section{Calculation of the critical exponents}

In this section \label{sec.critical.exponents}we will carry out the calculation
of the critical exponents associated with the transition to the Type A configuration,
which, as found in the previous section, is the only one that occurs in these
systems. As it is easy to see from the considerations in Section \ref{sec.model},
in order to obtain these exponents, we will need to include the effect of a
magnetic field applied to the system, which is what we do in the following.

We can make use of the results of the previous sections to consider the inclusion
of the applied magnetic field, \( \vec{h}=h\, \widehat{n}, \) in the direction
given by the normal vector \( \widehat{n} \), by redefining the conjugated
fields \( \vec{\xi }_{p\alpha } \) as\begin{equation}
\label{subs3}
\vec{\xi }_{p\alpha }\to \vec{\xi }_{p\alpha }+\left( \widehat{n}\cdot \widehat{n}_{\alpha }\right) h
\end{equation}
 for the \( p \)-spin cluster and\begin{equation}
\vec{\xi }_{1\alpha }\to \vec{\xi }_{1\alpha }+\left( \widehat{n}\cdot \widehat{n}_{\alpha }\right) h'
\end{equation}
 for the 1-spin cluster. The magnetic fields for both clusters scale as\begin{equation}
\label{h.scaling.relation}
h'=l^{y_{H}}h,
\end{equation}
 with \( y_{H} \) the magnetic critical exponent\cite{GOLDENFELD1994}.

Now, it is straightforward to see that the corresponding order parameter in
each sublattice for the 1-spin cluster is given by\begin{equation}
\label{1spin.op.with.field}
\vec{m}_{1\alpha }=\widehat{n}_{\alpha }\left\langle \tanh \left[ \widehat{n}_{\alpha }\cdot \left( \vec{\xi }_{1\alpha }+\widehat{n}h\right) \right] \right\rangle _{\mathcal{H}}.
\end{equation}
 For small values of \( h \), we can put\begin{equation}
\vec{m}_{1\alpha }\simeq \widehat{n}_{\alpha }\left\{ \left\langle \tanh \left[ \widehat{n}_{\alpha }\cdot \vec{\xi }_{1\alpha }\right] \right\rangle +\left\langle \sech ^{2}\left[ \widehat{n}_{\alpha }\cdot \vec{\xi }_{1\alpha }\right] \right\rangle \left( \widehat{n}\cdot \widehat{n}_{\alpha }\right) h+\mathcal{O}(h^{2})\right\} .
\end{equation}

\begin{figure}
{\par\centering \resizebox*{3in}{!}{\includegraphics{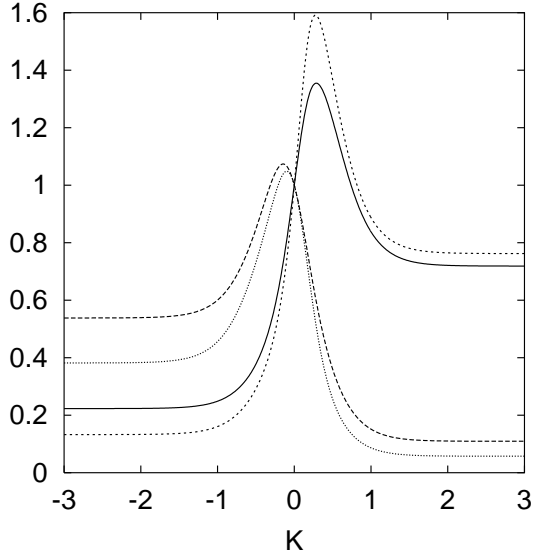}} \par}

\caption{\label{fig.ges}\protect\( \mathcal{K}\protect \) dependence of the functions
\protect\( G_{I}(\mathcal{K})\protect \) (solid line), \protect\( G_{SI}(\mathcal{K})\protect \)
(long dashed line) (given by \eqref{gising} and \eqref{gsi}, respectively),
\protect\( H_{I}(\mathcal{K})\protect \) (short dashed line), and \protect\( H_{SI}(\mathcal{K})\protect \)
(dotted line), (obtained from \eqref{hsi}, see text).}
\end{figure}

By making use of the exponential operator identity together with the van der
Waerden identity, we easily obtain, to the lowest order in the fields \( \vec{\xi } \)
and \( \vec{h} \) \begin{equation}
\label{prev.eq}
\vec{m}_{1\alpha }\simeq \left\{ A(\mathcal{K}')\sum _{\beta \neq \alpha }b_{1\beta }+F(\mathcal{K}')\left( \widehat{n}\cdot \widehat{n}_{\alpha }\right) h\right\} \widehat{n}_{\alpha },
\end{equation}
where \( A(\mathcal{K}') \) is given by \eqref{aes} and\begin{equation}
F(\mathcal{K}')=\rho _{x}^{z(p-1)}\left. \sech ^{2}(x)\right| _{x=0}.
\end{equation}
 For the \emph{kagomé} lattice (\( z=2 \) and \( p=3 \))\begin{equation}
\label{mag.kago}
F(\mathcal{K}')=\frac{1}{8}(3+4\sech ^{2}2\mathcal{K}'+\sech ^{2}4\mathcal{K}').
\end{equation}
 For the pyrochlore lattice (\( z=2 \) and \( p=4 \))\begin{equation}
\label{mag.pyro}
F(\mathcal{K})=\frac{1}{32}(10+15\sech ^{2}2\mathcal{K}+6\sech ^{2}4\mathcal{K}+\sech ^{2}6\mathcal{K}).
\end{equation}
 Let us now turn our attention to the 3-spin cluster in the \emph{kagomé} case.
We follow exactly the same procedure we have used to arrive to eq. \eqref{prev.eq},
but using the expression for the order parameter of this cluster, eq. \eqref{order.parameter.kagome}.
Thus, we first expand \( \vec{m}_{3\alpha } \) in powers of \( h \) and \( b_{3\alpha } \),
after the substitution \eqref{subs3} has been done, and make use of the differential
operator and van der Waerden identities. In doing so, we arrive at \begin{equation}
\vec{m}_{3\alpha }(b_{3A},b_{3B},b_{3C},\vec{h})\simeq \vec{m}_{3\alpha }(b_{3A},b_{3B},b_{3C})+\widehat{n}_{\alpha }(\vec{G}_{\alpha }(\mathcal{K})\cdot \vec{h})+\mathcal{O}(h^{2}).
\end{equation}
The first term on the right member of this expression is the same as \eqref{kagome.magnetization.zero.field},
whereas the second one is new, and the function \( \vec{G}_{\alpha }(\mathcal{K}) \)
is given by\begin{equation}
\vec{G}_{\alpha }(\mathcal{K})=\rho _{x}^{2}\rho _{y}^{2}\rho _{z}^{2}\left. \frac{\partial }{\partial \vec{h}}g_{\alpha }(x+\widehat{n}_{A}\cdot \vec{h},y+\widehat{n}_{B}\cdot \vec{h},z+\widehat{n}_{C}\cdot \vec{h})\right| _{\vec{h}=\vec{0}},
\end{equation}
 with \( g_{\alpha }(x,y,z) \) given as in \eqref{order3} and \eqref{ga.kagome}.
For arbitrary \( \widehat{n}_{A} \), \( \widehat{n}_{B} \), and \( \widehat{n}_{C} \),
\( \vec{G}_{\alpha } \) will depend on \( \alpha  \) in a complicated way.
However, in this work, we are only concerned about the cases studied in the
previous Section, namely, the standard Ising model, \( \widehat{n}_{A}=\widehat{n}_{B}=\widehat{n}_{C} \),
and the 2D spin ice problem on the \emph{kagomé} lattice, which verifies the
property \( \widehat{n}_{A}+\widehat{n}_{B}+\widehat{n}_{C}=\vec{0} \). In
these cases, it is easy to show that \begin{equation}
\vec{G}_{\alpha }(\mathcal{K})=G(\mathcal{K})\, \widehat{n}_{\alpha },
\end{equation}
 where the form of the function \( G(\mathcal{K}) \) does depend on whether
\( \widehat{n}_{A}=\widehat{n}_{B}=\widehat{n}_{C} \) or \( \widehat{n}_{A}+\widehat{n}_{B}+\widehat{n}_{C}=\vec{0} \).
In the first case (standard Ising model), this function is given by\begin{multline} 
\label{gising} 
G_{I}(\mathcal{K})=\frac{1}{192}\left( 138-\frac{320}{(1+e^{4\mathcal{K}})^{2}}+\frac{704}{1+e^{4\mathcal{K}}}-\frac{6144}{(3+e^{4\mathcal{K}})^{2}}+\frac{768}{3+e^{4\mathcal{K}}}+\frac{24(-7+3e^{4\mathcal{K}})}{(4-e^{4\mathcal{K}}+e^{8\mathcal{K}})^{2}}\right.\\ -\frac{8(-6+e^{4\mathcal{K}})}{4-e^{4\mathcal{K}}+e^{8\mathcal{K}}}-\frac{3(17+11e^{4\mathcal{K}})}{(1+e^{4\mathcal{K}}+2e^{8\mathcal{K}})^{2}}-\frac{3(1+80e^{4\mathcal{K}})}{1+e^{4\mathcal{K}}+2e^{8\mathcal{K}}}-\frac{48(-13-50e^{4\mathcal{K}}+63e^{8\mathcal{K}})}{(1+5e^{4\mathcal{K}}+e^{8\mathcal{K}}+e^{12\mathcal{K}})^{2}}\\ \left.+\frac{48(-13+18e^{4\mathcal{K}}+e^{8\mathcal{K}})}{1+5e^{4\mathcal{K}}+e^{8\mathcal{K}}+e^{12\mathcal{K}}}\right) , \end{multline}\\
whereas, in the second case (2D spin ice), it is given by\begin{multline} 
\label{gsi} 
G_{SI}(\mathcal{K})=\sech ^{2}2\mathcal{K}\left(533575+1162616\cosh 4\mathcal{K}+1014592\cosh 8\mathcal{K}+774088\cosh 12\mathcal{K}\right.\\ \left.+436732\cosh 16\mathcal{K}+178200\cosh 20\mathcal{K}+47360\cosh24 \mathcal{K}+6824\cosh 28\mathcal{K}+317\cosh 32\mathcal{K}\right)/\\ \left[1024(3\cosh 2\mathcal{K}+\cosh 6\mathcal{K}-2\sinh 2\mathcal{K})^{2}(-2\cosh 2\mathcal{K}+\sinh 2\mathcal{K})^{2}\right.\\ \left.\times (-1+5\cosh 4\mathcal{K}-3\sinh 4\mathcal{K})(1+3\cosh 4\mathcal{K}+\sinh 4\mathcal{K})^{2}\right]\\ -\sech ^{2}2\mathcal{K}\left(440942\sinh 4\mathcal{K}+661942\sinh 8\mathcal{K}+618862\sinh 12\mathcal{K}\right.\\ \left.+385138\sinh 16\mathcal{K}+163174\sinh 20\mathcal{K}+44430\sinh 24\mathcal{K}+6374\sinh 28\mathcal{K}+303\sinh 32\mathcal{K}\right)/\\ \left[1024(3\cosh 2\mathcal{K}+\cosh 6\mathcal{K}-2\sinh 2\mathcal{K})^{2}(-2\cosh 2\mathcal{K}+\sinh 2\mathcal{K})^{2}\right.\\ \left.\times(-1+5\cosh 4\mathcal{K}-3\sinh 4\mathcal{K})(1+3\cosh 4\mathcal{K}+\sinh 4\mathcal{K})^{2}\right]. 
\end{multline}\\
The \( \mathcal{K} \) dependence of these expressions has been depicted in
Fig.~\ref{fig.ges}.

The calculation for the pyrochlore lattice is identical to the one of the \emph{kagomé},
though the algebraic details are a little more involved. The final result of
the calculation is qualitatively the same as in the \emph{kagomé} case considered
above: the magnetization of the 4-spin cluster can be put as\begin{equation}
\vec{m}_{4\alpha }(b_{4A},b_{4B},b_{4C},b_{4D},\vec{h})\simeq \vec{m}_{4\alpha }(b_{4A},b_{4B},b_{4C},b_{4D})+H(\mathcal{K})\, (\widehat{n}_{\alpha }\cdot \vec{h})\, \widehat{n}_{\alpha }+\mathcal{O}(h^{2}).
\end{equation}
 where the first term on the right member is given by \eqref{order4} and \( H(\mathcal{K}) \)
is obtained from the \( g_{\alpha } \) function for the pyrochlore case as\begin{equation}
\label{hsi}
\vec{H}_{\alpha }(\mathcal{K})=\rho _{x}^{2}\rho _{y}^{2}\rho _{z}^{2}\left. \frac{\partial }{\partial \vec{h}}g_{\alpha }(x+\widehat{n}_{A}\cdot \vec{h},y+\widehat{n}_{B}\cdot \vec{h},z+\widehat{n}_{C}\cdot \vec{h})\right| _{\vec{h}=\vec{0}},
\end{equation}
 Again, for the cases considered in this work, \( \widehat{n}_{A}=\widehat{n}_{B}=\widehat{n}_{C}=\widehat{n}_{D} \)
(standard Ising) or \( \widehat{n}_{A}+\widehat{n}_{B}+\widehat{n}_{C}+\widehat{n}_{D}=\vec{0} \)
(spin ice), it can be easily shown that \( \vec{H}_{\alpha }(\mathcal{K})=H(\mathcal{K})\, \widehat{n}_{\alpha } \),
where the form of the function \( H(\mathcal{K}) \) does depend on whether
we are considering the standard Ising or the spin ice problem. We will represent
by \( H_{I}(\mathcal{K}) \) the first possibility and by \( H_{SI}(\mathcal{K}) \)
the second. The length of these expressions makes them of little practical use
and, therefore, we will not quote them here. It is enough for our purposes to
give their \( \mathcal{K} \) dependence by means of Fig.~\ref{fig.ges}. 

Now, by using the scaling relation \eqref{scaling}, we can equate the results
for the 1-spin cluster with those for the \( p \)-spin cluster, as we did in
Sec.~\ref{sec.critical.behavior}. In doing so, we obtained the following system
of equations for the internal fields\begin{equation}
\label{system.field}
\Phi (\mathcal{K})\, b_{p\alpha }+\left[ \Theta (\mathcal{K})-A(\mathcal{K}')\right] \sum _{\beta \neq \alpha }b_{p\beta }=\left[ F(\mathcal{K'})\, l^{-d+2y_{H}}-\Xi (\mathcal{K})\right] (\widehat{n}_{\alpha }\cdot \widehat{n})h,
\end{equation}
 where \( \Phi (\mathcal{K}) \), \( \Theta (\mathcal{K}) \), and \( A(\mathcal{K}) \)
are the same as in Sec.~\ref{sec.critical.behavior}, \( F(\mathcal{K}') \)
is given by \eqref{mag.kago} and \eqref{mag.pyro}, for the \emph{kagomé} and
pyrochlore lattices, respectively, and \( \Xi (\mathcal{K}) \) is given by
\( G(\mathcal{K}) \) (\( G_{I}(\mathcal{K}) \) for the standard Ising model,
and \( G_{SI}(\mathcal{K}) \) for the 2D spin ice) for the \emph{kagomé} lattice,
and \( H(\mathcal{K}) \) (\( H_{I}(\mathcal{K}) \) for the standard Ising
model, and \( H_{SI}(\mathcal{K}) \) for the spin ice) for the pyrochlore lattice.

In the limit case \( h=0 \), we recover the results obtained in Section \ref{sec.critical.behavior}
for the critical point. However, now we have an additional relation that allows
us to evaluate the magnetic critical exponent as\begin{equation}
y_{H}=\frac{1}{2}\left[ d+\frac{1}{\ln l}\ln \frac{\Xi (\mathcal{K}_{c})}{F(\mathcal{K}_{c})}\right] .
\end{equation}
 Moreover, the correlation length critical exponent is calculated by making
use of \eqref{thermal.eigenvalue} and \eqref{thermal.exponent} as\begin{equation}
\label{corre.exponent}
\nu =\frac{\ln l}{\ln \lambda _{T}},
\end{equation}
 where\begin{equation}
\lambda _{T}=\left. \left[ \frac{\partial }{\partial K}(\Phi (\mathcal{K})+(p-1)\Theta (\mathcal{K}))\right] \left[ (p-1)\frac{\partial }{\partial K'}A(\mathcal{K}')\right] ^{-1}\right| _{K_{c}}.
\end{equation}

The values of \( \nu  \) and \( y_{H} \) obtained for both standard Ising
spins and spin ice spins, and for both the \emph{kagomé} and the pyrochlores
lattices, are quoted in Table~\ref{table.critical.exponents}. The other exponents
for the different thermodynamic quantities can be obtained from these by making
use of the well known scaling relations\cite{GOLDENFELD1994} (see Table~\ref{tab.scaling.laws}).
The values listed as {}``exact{}'' are those obtained for the square and cubic
lattices, as they are expected to be the same as for the \emph{kagomé} and pyrochlore
lattices, respectively, on the basis of the universality hypothesis. The ones
corresponding to the square lattice are those obtained from the Onsager solution\cite{BAXTER1982},
whereas the ones for the cubic lattice are values obtained from high-precision
numerical studies\cite{LANDAU1994,BLOETE1995,TAPALOV1996}.
\begin{table}
{\centering \begin{tabular}{ccccc}
\hline 
\multicolumn{1}{c}{}&
\multicolumn{2}{c}{\emph{Kagomé} }&
\multicolumn{2}{c}{Pyrochlore }\\
\hline 
\multicolumn{1}{c}{}&
\( \nu  \)&
\( y_{H} \) &
\( \nu  \)&
\( y_{H} \)\\
\multicolumn{1}{l}{Standard Ising (\( \widehat{n}_{A}=\widehat{n}_{B}=\ldots  \))}&
\( 1.07 \)&
\( 1.64 \)&
\( 0.83 \)&
\( 2.32 \)\\
\multicolumn{1}{l}{Spin Ice (\( \sum _{\alpha }\widehat{n}_{\alpha }=\vec{0} \))}&
\( 1.07 \)&
\( 0.63 \)&
\( 0.83 \)&
\( 1.22 \)\\
\multicolumn{1}{l}{Exact (Standard Ising)\footnote{Refs.~\onlinecite{BAXTER1982,LANDAU1994,BLOETE1995,TAPALOV1996}}}&
\( 1 \)&
\( 1.875 \)&
\( 0.63 \)&
\( 2.48 \)\\
\hline 
\end{tabular}\par}

\caption{\label{table.critical.exponents}Values of the critical exponents.}
\end{table}

The first conclusion we can extract from the listed values for the standard
Ising problem is that the EFRG method provides very reliable estimates for the
critical exponents, in spite of the fact that we have chosen the smallest possible
clusters that include the effects of frustration. In this sense, the exponents
calculated in this work are some of the {}``best{}'' that can be found in
the literature, calculated in the framework of phenomenological RG approaches.
For example, the exponents obtained in our work are even better than the ones
obtained in the so called Mean Field RG method for Ising spins in the square
lattice, where clusters of 16 and 25 spins where used\cite{PLASCAK1999}. The
reason for this improvement is in the use of the van der Waerden identity, which
exactly takes into account the value of the autocorrelation function. It is
also important to note that the estimation of \( \nu  \) for the \emph{kagomé}
lattice is better than the one for the pyrochlore lattice. The reason is that
the scaling factor we have used, \( l=(N/N')^{1/d}=p^{1/d} \), is smaller for
the 3D lattice than for the 2D one. It has been noted by some authors, that
the values of the critical exponents calculated with this method can be improved
by using an alternative definition of the size of the clusters, in terms of
the number of interactions including the ones with the spins creating the effective
field\cite{SLOTTE1987}. This point will be further investigated in a future
work\footnote{%
Another possible definition of the scaling factor we can use consists in calculating
from \eqref{corre.exponent} the value of $l$ that leads to the exact value of
$y_t$ for the standard Ising model, and make use of this value to estimate the
corresponding magnetic critical exponent. The values of $l$ obtained in this
way are $l=1.67$ and $l=1.42$ for the \emph{kagomé} and pyrochlore lattices,
respectively. The corresponding magnetic critical exponents are $y_H=1.68$ and
$y_H=2.58$, for the standard Ising model in the \emph{kagomé} and pyrochlore
lattices, respectively, and $y_H=0.60$ and $y_H=1.13$, for the 2D spin ice and
the 3D spin ice, respectively. 
}. 

The last important conclusion we can extract from Table~\ref{table.critical.exponents}
is that the magnetic critical exponent and, thus, all the exponents depending
on this one (\( \beta  \), \( \gamma  \), \( \delta  \), and \( \eta  \)),
are \emph{different for the spin ice case}. If we take into account the good
agreement found for the standard Ising case, we can expect the estimations of
the critical exponents for the spin ice case to be of the same quality. Actually,
if we assume that the error in the determination of the critical exponents for
the spin ice case is the same as in the standard Ising case (\( \Delta y_{H}/y_{H}=0.13 \)
for the \emph{kagomé} Ising case and \( \Delta y_{H}/y_{H}=0.06 \) for the
pyrochlore Ising), and we see no reason for this not to be the case, we can
put an upper and lower bounds for the real value of this exponent, \( y_{H}=0.63\pm 0.08 \)
for the \emph{kagomé} spin ice, and \( y_{H}=1.22\pm 0.07 \) for the pyrochlore
spin ice. Apart from a possible relative error of about a 10\% in the worse
case, they are obviously very different from the \( y_{H} \) of the standard
Ising model and, in fact, we have not found in the literature any reference
to any other system with these values for \( y_{H} \). A \emph{naive} interpretation
of these exponents could make us think that our results violate the universality
hypothesis. However, it is easy to realize that this is not the case. Indeed,
the universality hypothesis states that the critical exponents are the same
for different systems with Hamiltonians which internal symmetries are the same,
independently of details as the microscopic origin of the interactions or the
geometry of the lattice. The results we found for the standard Ising model in
both the \emph{kagomé} and pyrochlore lattices are in good agreement with this
idea. Moreover, the fact that the thermal exponent does not change when going
from Ising spins with a single anisotropy axis, to the spin ice problem, is
also in agreement with the universality hypothesis, as we have previously shown
that the spin ice problem maps on an Ising Hamiltonian with effective interactions
given by \( \mathcal{K}=K\, \cos \theta  \). However, when we include the effect
of an uniform magnetic field in the spin ice case, the Zeeman term cannot be
mapped to the Zeeman term in the Ising Hamiltonian with a single anisotropy
axis, which is the same for all the spins, but we can see it as a Zeeman which
is different for spins located in different sublattices, i.e., it is equivalent
to some kind of \emph{staggered} field subject to the condition \( \sum _{\alpha }\vec{h}_{\alpha }=\vec{0} \),
where by \( \vec{h}_{\alpha } \) we represent the magnetic field acting on
each sublattice. Certainly, this Zeeman term has not the same symmetry as the
original Zeeman term of the Ising spin model with a single anisotropy axis and,
therefore, \emph{we should expect the corresponding magnetic critical exponent
to be different from the one of the standard Ising model.} Of course, it would
be extremely important to have experimental confirmation of this fact, or even
estimations of the magnetic critical exponent obtained from high precision Monte
Carlo simulations but, as far as we know, this is the first time that this result
is reported.
\begin{table}
{\centering \begin{tabular}{c|c}
\hline 
Critical exponent&
Scaling relation\\
\hline 
\hline 
\( \alpha  \)&
\( 2-d/y_{H} \)\\
\hline 
\( \beta  \)&
\( (d-y_{H})/y_{T} \)\\
\hline 
\( \gamma  \)&
\( (2y_{H}-d)/y_{T} \)\\
\hline 
\( \delta  \)&
\( y_{H}/(d-y_{H}) \)\\
\hline 
\( \eta  \)&
\( 2-2y_{H}+d \)\\
\hline 
\( \nu  \)&
\( 1/y_{t} \)\\
\hline 
\end{tabular}\par}

\caption{\label{tab.scaling.laws}Expressions of the critical exponents in terms of
the RG critical exponents \protect\( y_{T}\protect \) and \protect\( y_{H}\protect \).}
\end{table}

Another interesting point we want to notice here is the connection of these
spin ice-like systems with the well known vertex models\cite{BRAMWELL1998,BAXTER1982}.
In fact, it is easy to see that the 2D spin ice of this work is related to the
8-vertex model, whereas the spin ice on the pyrochlore is related to the 16-vertex
model. On the other hand, it is well known that these vertex models have the
peculiarity that the critical exponents depend continuously on the value of
the exchange coupling, and this is precisely what we would expect in our model
if we allow for arbitrary orientations of the unit vectors \( \widehat{n}_{\alpha } \),
for which \( \widehat{n}_{A}\cdot \widehat{n}_{B}\neq \widehat{n}_{A}\cdot \widehat{n}_{B}\neq \ldots  \).
We think that clearly establishing this connection, though outside of the scope
of this work, deserves further consideration, as it seems that these spin ice-like
systems are one of the few physical realizations of these vertex models\footnote{%
Even though this is only the case for very special cases, as the ones considered
in this work, because spin ice systems with arbitrary unitary vectors are, in
general, incompatible with the symmetries of the crystalline lattice.
}. In this sense, we think that the present RG scheme could prove to be useful
in obtaining qualitative information about critical properties of the 16-vertex
model, about which very little is known.

\section{The GCC method applied to the Spin Ice problem}

As stated in the\label{sec.gcc.method} Introduction, it is interesting to compare
the results obtained in the previous sections with those obtained in the framework
of the GCC method, which has been successfully applied to the problem of Heisenberg
frustrated lattices by the present authors\cite{GARCIA-ADEVA2000a,GARCIA-ADEVA2000b}.
In the GCC framework, we initially consider isolated units (tetrahedra or triangles,
for the pyrochlore and \emph{kagomé} lattices, respectively), and later add
the interactions with the surrounding units as unknown effective fields to be
determined by a self consistent condition, which is obtained by comparing the
order parameter in the cluster with that of the isolated ion\cite{ELLIOTT1960,GARCIA-ADEVA2000a}.
We will analyze here the case in which the applied magnetic field is zero, even
though the generalization is straightforward.

As in the previous sections, we will characterize each spin in the frustrated
unit by a different order parameter. Therefore, the Hamiltonian of the cluster
formed by one Ising spin is given by\begin{equation}
\mathcal{H}_{1\alpha }=2K\, \vec{s}_{1\alpha }\sum _{\beta \neq \alpha }\vec{h}_{\beta }=2\vec{s}_{1\alpha }\cdot \vec{\xi }_{1\alpha },
\end{equation}
 where \( \vec{h}_{\beta } \) are the internal fields created by the neighboring
spins outside the unit, and point along the directions given by \( \widehat{n}_{\beta } \),
and we have taken into account that there are 2 neighbors in each sublattice.
The Hamiltonian of the cluster formed by \( p \) spins can be put as\begin{equation}
\mathcal{H}_{p}=K\sum _{\alpha \neq \beta }\vec{s}_{1\alpha }\cdot \vec{s}_{1\beta }+\sum _{\alpha }\vec{s}_{1\alpha }\cdot \vec{\xi }_{p\alpha }.
\end{equation}
 We can then see that the corresponding partition function of the unit has the
same functional form as eqs. \eqref{order.parameter.kagome} and \eqref{order.full}
for the \emph{kagomé} and pyrochlore lattices, respectively, except for the
fact that the average with respect to the full Hamiltonian is absent, and the
conjugated fields \( \vec{\xi }_{\alpha } \) appearing in \( \Pi _{\alpha } \)
depend on the internal fields \( \vec{h}_{\beta } \), rather than on the spin
variables of the surrounding spins outside the cluster. As stated above, these
internal fields are evaluated by imposing the self consistent condition of equating
the magnetization per spin in the internal field with that of a unit in the
internal field, which can be mathematically stated as\begin{equation}
\vec{m}_{1\alpha }\left( \vec{h}_{A},\vec{h}_{B},\vec{h}_{C},\ldots \right) =\vec{m}_{p\alpha }\left( \vec{h}_{A},\vec{h}_{B},\vec{h}_{C},\ldots \right) .
\end{equation}
 Obviously, this equation can only be solved numerically in the general case.
However, near the critical point, we expect the internal fields to be very small,
so we can expand the order parameters to first order in the internal fields.
In doing so, we obtain\begin{equation}
m_{1\alpha }\simeq 2\mathcal{K}\sum _{\beta \neq \alpha }h_{\beta },
\end{equation}
\begin{equation}
m_{p\alpha }=2\mathcal{K}\left( A_{1}(\mathcal{K})\, h_{\alpha }+A_{2}(\mathcal{K})\, \sum _{\beta \neq \alpha }h_{\beta }\right) ,
\end{equation}
 where \( \mathcal{K}=K\cos \theta  \), as before, and \[
A_{1}(\mathcal{K})=\left\{ \begin{array}{l}
\frac{1-e^{-4\mathcal{K}}}{1+3e^{-4\mathcal{K}}}\\
\frac{3(1-e^{-8\mathcal{K}})}{1+4e^{-6\mathcal{K}}+3e^{-8\mathcal{K}}}
\end{array}\right. ,\]
 and\begin{equation}
A_{2}(\mathcal{K})=\left\{ \begin{array}{l}
\frac{1+e^{-4\mathcal{K}}}{1+3e^{-4\mathcal{K}}}\\
\frac{3+4e^{-6\mathcal{K}}+e^{-8\mathcal{K}}}{1+4e^{-6\mathcal{K}}+3e^{-8\mathcal{K}}}
\end{array}\right. ,
\end{equation}
 where the upper (lower) expression corresponds to the \emph{kagomé} (pyrochlore)
lattice.

Then, we arrive to the following system of equations for the effective fields\begin{equation}
\Phi (\mathcal{K})\, h_{\alpha }+\Theta (\mathcal{K})\, (h_{\beta }+h_{\gamma }+\ldots )=0,\, \, \, \, (\alpha \neq \beta \neq \gamma \neq \ldots ),
\end{equation}
 where\begin{equation}
\Phi (\mathcal{K})=A_{1}(\mathcal{K}),
\end{equation}
 and\begin{equation}
\Theta (\mathcal{K})=A_{2}(\mathcal{K})-\left\{ \begin{array}{l}
1\\
2
\end{array}\right\} =\left\{ \begin{array}{l}
-\frac{2}{3+e^{4\mathcal{K}}}\\
\frac{1-4e^{-6\mathcal{K}}-5e^{-8\mathcal{K}}}{1+4e^{-6\mathcal{K}}+3e^{-8\mathcal{K}}}
\end{array}\right. .
\end{equation}
 Again, the upper (lower) expression corresponds to the \emph{kagomé} (pyrochlore)
lattice.

This system of equations has two non-trivial solutions for\begin{equation}
\Phi (\mathcal{K})=\Theta (\mathcal{K})
\end{equation}
 and \begin{equation}
\Phi (\mathcal{K})=-(p-1)\Theta (\mathcal{K}).
\end{equation}
 The first one corresponds to a configuration in which \( \sum _{\alpha }h_{\alpha }=0 \),
whereas the second one corresponds to a configuration in which \( h_{A}=h_{B}=h_{C}=\ldots  \),
that is, they are the same configurations we found in the EFRG formulation.
It is easy to verify that the first kind of solution never occurs for any finite
value of \( \mathcal{K} \), whereas the second one occurs for a value of \( \mathcal{K}_{c}=0.402 \)
for the \emph{kagomé} lattice and \( \mathcal{K}_{c}=0.227 \) for the pyrochlore
lattice, in very good agreement with our previous results (see Table~\ref{table1}).

\section{Connection between the GCC and EFRG schemes}

From the previous section\label{sec.connection}, it seems clear that both approaches
presented in this work are somehow related. In this section we will furnish
this connection. We will consider a system with only one vectorial order parameter.
The generalization to an arbitrary number of order parameters being straightforward.

The Hamiltonian of the one spin cluster with only NN interactions can be put
as\begin{equation}
\mathcal{H}=\vec{s}_{1}\cdot \vec{\xi }_{1}=\vec{s}_{1}\cdot \left( K\sum _{i}^{z}\vec{s}_{i}\right) ,
\end{equation}
 where \( z \) is the number of NN. The corresponding order parameter can be
obtained then from the Callen-Suzuki identity and, in general, it will be a
function of the conjugated field \( \vec{\xi }_{1} \)\begin{equation}
\vec{m}_{1}=\left\langle \frac{\partial }{\partial \vec{\xi }_{1}}\ln Z_{1}(\vec{\xi }_{1})\right\rangle _{\mathcal{H}}=\left\langle \vec{F}(\vec{\xi }_{1})\right\rangle _{\mathcal{H}}=\left\langle \sum _{\mu }F_{\mu }\left( K\sum _{i}^{z}\vec{s}_{i}\right) \, \widehat{n}_{\mu }\right\rangle _{\mathcal{H}},
\end{equation}
 where \( \widehat{n}_{\mu } \) define the coordinate system. Obviously, we
can expand the function \( \vec{F} \) in powers of \( \vec{s}_{i} \) as \begin{multline} \vec{m}_{1}=\sum _{\mu }\widehat{n}_{_{\mu }}\left\langle F_{\mu }(\vec{0})+\sum _{i}^{z}\left. \frac{\partial F_{\mu }}{\partial \vec{s}_{i}}\right| _{\vec{s}_{i}=\vec{0}}\cdot \vec{s}_{i}+\frac{1}{2!}\sum _{i,j}\left. \frac{\partial ^{2}F_{\mu }}{\partial \vec{s}_{i}\cdot \partial \vec{s}_{j}}\right| _{\vec{s}_{i}=\vec{s}_{j}=\vec{0}}\vec{s}_{i}\cdot \vec{s}_{j}+\ldots \right\rangle _{\mathcal{H}}\\ =\sum _{\mu }\widehat{n}_{_{\mu }}\left[ F_{\mu }(\vec{0})+\sum _{i}^{z}\left. \frac{\partial F_{\mu }}{\partial \vec{s}_{i}}\right| _{\vec{s}_{i}=\vec{0}}\cdot \left\langle \vec{s}_{i}\right\rangle _{\mathcal{H}}+\frac{1}{2!}\sum _{i,j}\left. \frac{\partial ^{2}F_{\mu }}{\partial \vec{s}_{i}\cdot \partial \vec{s}_{j}}\right| _{\vec{s}_{i}=\vec{s}_{j}=\vec{0}}\left\langle \vec{s}_{i}\cdot \vec{s}_{j}\right\rangle _{\mathcal{H}}+\ldots \right] .
\end{multline}

If we now apply the decoupling approximation to this expression\begin{equation}
\left\langle s_{i}^{\mu }s_{j}^{\nu }\cdots s_{k}^{\eta }\right\rangle _{\mathcal{H}}\simeq \left\langle s_{i}^{\mu }\right\rangle _{\mathcal{H}}\left\langle s_{j}^{\nu }\right\rangle _{\mathcal{H}}\cdots \left\langle s_{k}^{\eta }\right\rangle _{\mathcal{H}}=h_{1}^{\mu }\, h_{1}^{\nu }\cdots h_{1}^{\eta },
\end{equation}
 where \( \vec{h}_{1}=\left\langle \vec{s}_{i}\right\rangle _{\mathcal{H}} \),
and take into account that the variables in the derivatives are dummy variables,
that is, we can put\begin{equation}
\left. \frac{\partial F_{\mu }(\vec{s}_{i})}{\partial \vec{s}_{i}}\right| _{\vec{s}_{i}=\vec{0}}=\left. \frac{\partial F_{\mu }(\left\langle \vec{s}_{i}\right\rangle )}{\partial \left\langle \vec{s}_{i}\right\rangle }\right| _{\left\langle \vec{s}_{i}\right\rangle =\vec{0}},
\end{equation}
 the series can be resumed in terms of the effective internal field \( \vec{h}_{1} \)
to give\begin{equation}
\vec{m}_{1}\simeq \vec{F}(z\, K\, \vec{h}_{1}).
\end{equation}
 Obviously, we can do the same with the \( p \) cluster order parameter, and
the result is\begin{equation}
\vec{m}_{p}\simeq \vec{G}((z-1)\, K\, \vec{h}_{p}),
\end{equation}
 where \( \vec{h}_{p}=\left\langle \vec{s}_{i}\right\rangle _{\mathcal{H}} \)
is the effective internal field obtained from the \( p \)-spin cluster. We
can then easily see that the GCC self consistency condition corresponds to the
particular case of the finite size scaling hypothesis in which one assumes\begin{equation}
\vec{m}_{p'}=\vec{m}_{p},
\end{equation}
 and\begin{equation}
\vec{h}_{p'}=\vec{h}_{p}.
\end{equation}

Thus, the GCC model corresponds to the lowest order of the EFRG method, in which
correlations outside the cluster are neglected and the scaling relation is substituted
by strict equality.

Therefore, the value of the critical point obtained in both methods, to this
order of approximation, is the same, as it does not depend on the particular
scaling coefficient. However, this has profound consequences for the calculation
of other quantities, as the critical exponents, which do depend on the particular
scaling relation, as have been noted by some authors\cite{PLASCAK1999}. The
differences between the values listed in Table~\ref{table1} are not related
to the use of the EFRG or GCC methods, but to the fact that we have used the
van der Waerden identity in the case of the EFRG. However, we could make use
of this identity in the GCC approach, at least for Ising spins, and the results
would be identical to this order of approximation (i.e. up to terms linear in
\( K_{c} \)).

It is also important to stress that the results of this section are not limited
to the connection between the EFRG and GCC methods applied to GFAF, as we have
not made any assumption about the sizes of the clusters. In this sense, this
connection also holds for the original formulation of both the Constant Coupling
and EFRG methods, in which clusters with 1 and 2 spins are compared\cite{ELLIOTT1960}.
In other words, the Constant Coupling method (in whatever formulation) is the
lowest order approximation of the EFRG method, corresponding to neglecting multispin
correlations and assuming that the anomalous dimension of the order parameter
is zero.

\section{Conclusions}

In this work \label{sec.conclusions}we have studied the critical behavior of
ferro- and antiferromagnetically coupled Ising spins with local anisotropy axes
in geometrically frustrated geometries (\emph{kagomé} and pyrochlores lattices).
The present problem, applied to the pyrochlore lattice with local anisotropy
axes in high symmetry \( \left\langle 1\, 1\, 1\right\rangle  \)-like directions
corresponds to the spin ice problem. The critical points have been calculated
by means of the effective field renormalization group technique, adequately
generalized to deal with the non-Bravais character of the lattices, and by the
generalized constant coupling approach. These methods allow us to evaluate both
the critical points, if any, and the corresponding ground state configurations.
The results obtained are in very good agreement with exact results known for
the Ising Hamiltonian in the 2D \emph{kagomé} lattice and Monte Carlo calculations
for the pyrochlore lattice. For ferromagnetic interactions, geometrical frustration
prevents the formation of long range order, whereas for antiferromagnetic ones,
there is a transition at a finite temperature to a long range ordered state
in which all the spins of the unit point outside or inside the frustrated unit.
These results can be traced back to the fact that this spin ice problem can
be mapped to the standard Ising model with an effective interaction given by
\( \mathcal{K}=K\cos \theta  \), where \( \theta  \) is the angle between
any 2 anisotropy directions. Therefore, ferromagnetic interactions in the spin
ice problem correspond to effective antiferromagnetic ones, and vice versa.
Incidentally, the connection between the effective field renormalization group
technique and the generalized constant coupling method is furnished. It is shown
that the GCC method corresponds to the lowest approximation of the EFRG scheme,
in which correlations outside the cluster are neglected and the anomalous dimension
of the order parameter is taken to be zero.

In addition, the critical exponents of these systems have been calculated for
the transition to the long range ordered state that occurs for effective ferromagnetic
interactions. The calculated values of the critical exponents for spins with
parallel anisotropy axes (the standard Ising model), are found to be in very
good agreement with both those obtained from exact results for the 2D square
lattice and those calculated computationally for the cubic lattice, in spite
of the small size of the clusters considered in this work. This fact is in agreement
with the universality hypothesis. For the spin ice problem (in both the \emph{kagomé}
and the pyrochlore lattice), the thermal exponent has the same value as for
standard Ising spins, as both Hamiltonians can be identically mapped on to each
other. However, the magnetic exponent is found to be different from that of
the standard Ising model in the presence of an uniform magnetic field. This
result is expected, as the Zeeman term in this case has different symmetry properties
than in the standard Ising model. In any case, independent confirmation of this
fact, by experimental or Monte Carlo simulations, would be desirable.

We think that the techniques presented in this work will prove to be useful
in studying other phenomena present in the spin ice system. For example, it
would be interesting to study how the inclusion of long range dipolar interactions
modify the present results, that is, if these additional interactions could
remove the degeneracy of the ground state so a long range ordered state is formed
at any finite temperature for the ferromagnetic case. So far, there are numerical results and mean field calculations which indicate that the degeneracy of the ground state is weakly lifted by long range dipolar interactions, and there is a finite temperature much lower than the energy scale defined by the dipolar interactions at which a long range ordered state given by the ice rules is formed \cite{GINGRAS2000b,MELKO2000}, but we think
it would be important also to support this result with analytical calculations that go beyond mean field theory. Also, the effect of a small amount of dilution in the crystalline lattice is
an intriguing problem which can be dealt with in the present formalisms. Moreover,
even though the extension is not trivial, the present EFRG scheme can be generalized
to deal with Heisenberg spins and, thus, shed some light in the fascinating
phenomena of geometrical frustration.

Of course, the EFRG scheme is not exempt of limitations. For example, the inclusion
of fluctuating corrections coming from correlations further than NN interactions
is not an easy task. However, there are some methods to deal with this problem
\cite{JIANG1996}, though they are highly technical and outside of the scope
of the present work. This, and some of the problems mentioned in the above paragraph
are the directions of our present research.

\begin{acknowledgments}
AJGA wishes to thank the Spanish MEC for financial support under the Subprograma
General de Perfeccionamiento de Doctores y Tecnólogos en el Extranjero.
\end{acknowledgments}

\end{document}